\documentclass[12pt]{article}
\pdfoutput=1
\usepackage{amsfonts,amsthm}
\usepackage{amsmath,amssymb}
\usepackage{hyperref}
\usepackage{srcltx}
\usepackage{tikz}
\newcommand{\be}{\begin{equation}}
\newcommand{\ee}{\end{equation}}
\newcommand{\bea}{\begin{eqnarray}}
\newcommand{\eea}{\end{eqnarray}}

\renewcommand{\Re}{\mathrm{Re }}
\renewcommand{\Im}{\mathrm{Im }}

\newcommand{\lr}[1]{ \langle #1 \rangle}
\newcommand{\Z}{\mathbb{Z}}

\providecommand{\RR}{{\mathbb{R}}}

\providecommand{\mtrx}[1]{\begin{pmatrix} #1 \end{pmatrix}}

\providecommand{\bs}[1]{\boldsymbol{#1}}

\def\lsim{\mathrel{\rlap{\lower4pt\hbox{\hskip1pt$\sim$}}
    \raise1pt\hbox{$<$}}}         
\def\gsim{\mathrel{\rlap{\lower4pt\hbox{\hskip1pt$\sim$}}
    \raise1pt\hbox{$>$}}}         

\renewcommand{\Z}{\mathbb{Z}}

\title{Symmetry breaking patterns in 3HDM}

\author{I.~P.~Ivanov$^{1,2}$, C.~C.~Nishi$^{3,4}$  
\\
  {\small $^1$ Ghent University, Department  of Physics and Astronomy, Proeftuinstraat 86, 9000, Gent, Belgium}\\
  {\small $^2$ IFPA, Universit\'{e} de Li\`{e}ge, All\'{e}e du 6 Ao\^{u}t 17, b\^{a}timent B5a, 4000, Li\`{e}ge, Belgium}\\
  {\small $^3$ Universidade Federal do ABC - UFABC, 09.210-170, Santo Andr\'e, SP, 
Brasil}\\
  {\small $^4$ Maryland Center for Fundamental Physics, 
University of Maryland, College Park, MD 20742, USA}
  }

\begin{document}

\maketitle

\bigskip
\begin{abstract}
An attractive feature of New Physics models with multiple Higgs fields is 
that they are equipped with discrete symmetry groups in the Higgs and flavour sectors.
These symmetry groups are often broken at the global minimum of the Higgs potential,
either completely or to a proper subgroup, with certain phenomenological consequences.
Here, we systematically explore these symmetry breaking patterns in the scalar sector 
of the three-Higgs-doublet model (3HDM).
We use the full list of discrete symmetry groups allowed in 3HDM, and
for each group we find all possible ways it can break by the Higgs vacuum expectation value alignment.
We also discuss the interplay between these symmetry groups and various forms of $CP$-violation 
in the scalar sector of 3HDM. 
Not only do our results solve the problem for 3HDM, but they also hint at several 
general features in multi-scalar sectors.
\end{abstract}



\section{Introduction}

Beyond the Standard Model (bSM) constructions often use multiple Higgs fields or 
other scalars \cite{CPNSh,inert,neutrino-reviews}.
A large number of scalar fields with equal quantum numbers allows one
to equip these models with extra global symmetries such as Higgs-family,
generalized-$CP$, or flavour symmetries.
Existence of these symmetries and their spontaneous breaking upon minimization of the scalar
potential has strong impact on phenomenology in the scalar and flavour sectors, 
as well as astroparticle consequences.

Any bSM model built on an extended scalar sector --- especially with multiple Higgs 
doublets --- must also specify how scalars interact with fermions
in order to be complete and to claim its relevance to the experiment.
This issue requires much care. Generic Yukawa couplings between different scalar doublets and the fermions
will lead to unacceptably large flavour-changing neutral currents (FCNC) and violate 
electroweak precision and flavour physics constraints. 
A popular way to naturally suppress FCNC is to impose discrete flavour-blind symmetries
not only on the scalar but also on Yukawa sector of the theory \cite{nfc}.
The two-Higgs-doublet model (2HDM) \cite{2HDMreview} with its four types of the $\Z_2$ symmetry
in the Yukawa sector is one well-known example 
of the interplay between the scalar sector and flavour observables via discrete symmetries.
With several Higgs doublets, one has more freedom in imposing discrete symmetries
both on the scalar potential and on the Yukawa sector.
One key result here is that if quark masses and mixing are supposed to come
from coupling to several active Higgs doublets via symmetry-related Yukawa textures, 
only complete breaking of the flavour symmetry by the Higgs 
potential can lead to a viable quark masses and the CKM matrix
\cite{NHDM-quarks-general}.

To balance previous arguments, we remark that certain 
multi-doublet models can easily avoid restrictions coming from flavour observables.
For example, one can assume that new scalars are fermiophobic, so that 
the bSM sector of the model is limited to interaction between various scalars.
The model can then exhibit a very SM-like collider phenomenology, 
possibly with non-trivial scalar dark matter implications.
A simple example of such a situation is given by models with one additional inert doublet \cite{inert}.
Another opportunity, driven by the fact the newly found boson seems to 
have the SM Yukawa couplings to the third family of fermions, is that additional Higgs 
states which couple to ligher fermions are much heavier.
What are the scalar mass eigenstates, how they interact with each other,
does the interactions stabilize some of these states:
all these questions rely on the symmetry breaking patterns.

We also want to stress that very similar issues arise in flavour models,
which make use not of several Higgs doublets but of multiple electroweak singlet 
fields (flavons) which carry flavour charges \cite{neutrino-reviews}.
In general, such a flavour symmetry must have some (irreducible or not) 
three-dimensional representation to accommodate the three fermion families.
In this case the Higgs sector can be that of the SM or its minimal supersymmetric 
extension, and flavour symmetry breaking is communicated to the SM through higher 
order operators.
The flavour symmetry is then expected to be broken at much higher energies 
than when Higgs doublets carry flavour. FCNC effects through flavon exchange are 
also expected to be very much suppressed because of their large masses and 
observable flavour violating effects, when it exists, should be induced from other 
sectors of the theory lying at intermediate energy scales.
This is another instance where the patterns of symmetry breaking by a scalar 
potential are crucial.

All these examples underline the important role played by (discrete) symmetries in 
various multi-scalar models, 
which necessitates their systematic investigation in each class of models. 
In this work, we report the symmetry breaking analysis for the three-Higgs-doublet 
model (3HDM). 
Although we will use the notation and nomenclature of the 3HDM, 
the results we obtain are relevant not only to the 3HDM {\em per se},
but also to other models with three scalar fields 
carrying the same SM quantum numbers and additional conserved $U(1)$ charges.

Multi-Higgs-doublet models represent a rather conservative class of bSM models which 
has several remarkable
phenomenological consequences for $CP$-violation 
\cite{weinberg3HDM,branco1980,geometricCP,DeshpandeHe1994,CPbook}
and in the scalar, flavour, and neutrino sectors \cite{3HDM-S3,3HDM-A4,NHDMs-recent,Nishi2007,NHDM-quarks}.
The full list of discrete symmetry groups allowed in the 3HDM scalar sector was presented recently in 
\cite{classification3HDM}, and an efficient geometric method suitable for minimization of highly symmetric
potentials was developed in \cite{minimization2012}.
Building on these results, we present an exhaustive case by case investigation of how each of the allowed
discrete symmetry group in 3HDM can break by vacuum expectation values (vev) alignments.

The purpose of this study is two-fold.
First, 3HDMs with various symmetries are often used in bSM model-building.
Although some specific breaking patterns for some groups have already been explored previously,
we present the first exhaustive list of all possibilities offered in pure 3HDMs (that is, 
three Higgs doublets without
any extra scalars) with renormalizable potentials.
Second, this exhaustive list hints at certain tendencies, which might hold for models with $N$ Higgs doublets (NHDM)
or even for more elaborate Higgs sectors.
Thus, this work represents a step towards establishing general properties of discrete symmetry breaking patterns
in multi-scalar models.
\\

The structure of this paper is the following.
In the next section we provide the context for our study by discussing relevant 
symmetry-related results in multi-Higgs-doublet models.
Sections~\ref{section-main1} and \ref{section-main2} contains detailed analyses 
of symmetry breaking options available for each discrete group in 3HDM, without and with triplet irreducible representations.
In Section~\ref{section-summary} we summarize the emerging picture,
discuss its implications for more complicated Higgs sector, and draw conclusions.
Mathematical details on minimization of the Higgs potential for certain symmetry groups
are given in Appendices.

\section{Overview of symmetry-related results}\label{section-overview}

\subsection{Preliminary technical remarks}

In order to avoid possible misunderstanding, let us start with two technical but important remarks 
on what kind of symmetry groups we consider in the NHDM scalar sector.

First, when we work with $N$ Higgs doublets $\phi_i$ with identical quantum numbers, we 
can perform a unitary (Higgs family) or anti-unitary (generalized $CP$, GCP) 
global transformation in the $N$-dimensional space
of Higgs doublets: $\phi_i \mapsto U_{ij}\phi_j$ or $\phi_i \mapsto U_{ij}\phi_j^*$,
with $U_{ij} \in U(N)$, see more details and references in \cite{2HDMreview}. 
These transformations respect electroweak symmetry and bring a chosen potential to another viable potential.
If such a transformation leaves the potential invariant, we say that we have a symmetry.
By construction, any Higgs potential is symmetric under the simultaneous and equal rephasings
of all doublets, $\phi_i \mapsto e^{i\alpha}\phi_i$, which form the group $U(1)$.
We are interested not in this trivial symmetry but in {\em additional symmetries}
which some potentials can also have. When we talk about non-trivial symmetries of the potential,
we mean symmetries up to this overall rephasing. Technically, we search for symmetry groups $G$ 
which are subgroups not of $U(N)$, and not even of 
$SU(N)$, but of $PSU(N) \simeq U(N)/U(1) \simeq SU(N)/\Z_N$.

Second, when we say that a potential has symmetry group $G$, we mean that $G$ contains
{\em all} symmetry content of the given potential. 
This is somewhat different from the usual
approach when one just imposes a symmetry group on the model. 
Here, we additionally check that there is no other symmetry transformation
which could possibly arise. All the groups we mention below pass this check.
In terminology of \cite{classification3HDM,abelianNHDM}, we are interested only in realizable groups.
It also means that the Higgs doublets are always assumed to be in a faithful (but not necessarily irreducible)
representation of the group $G$.

\subsection{Symmetries in 3HDM}

With these remarks in mind, let us summarize the symmetry results
in the scalar sector of the two-Higgs-doublet model (2HDM), focusing on discrete symmetry groups
(for a more detailed exposition, see \cite{2HDMreview} and reference therein).
The 2HDM scalar potential can have only three discrete realizable symmetry groups:
$\Z_2^*$, $\Z_2\times\Z_2^*$, or $(\Z_2)^2\times\Z_2^*$, where 
$\Z_2^*$ denotes a GCP symmetry.
In other words, it is impossible to have an explicitly $CP$-violating 2HDM potential 
with some Higgs family symmetry.
Minimization of the Higgs potential leads to a vev alignment which either keeps
the symmetry intact or removes just one $\Z_2$-factor: 
$(\Z_2)^k \to (\Z_2)^{k-1}$, $k=1,2,3$, including GCP symmetries
\cite{Ivanov2HDMMinkowski}.
This result already illustrates the important feature that sufficiently large symmetry groups 
cannot be broken completely.

\tikzset{node distance=1.2cm, auto}
\begin{figure}[!htb]
   \centering
\begin{tikzpicture}
  \node (E) {$\{e\}$};
  \node (Z2) [above of=E] {$\Z_2$};
  \node (Z2Z2) [above of=Z2] {$\Z_2\times\Z_2$};
  \node (Z3) [node distance=2.5cm, right of=Z2Z2] {$\Z_3$};
  \node (Z4) [node distance=2.5cm, left of=Z2Z2] {$\underline{\Z_4}$};
  \node (D8) [above of=Z4] {$\underline{D_4}$};
  \node (A4) [above of=Z2Z2] {$\underline{A_4}$};
  \node (S4) [node distance = 1.8cm, above of=A4] {$\underline{S_4}$};
  \node (D6)[node distance=2.5cm, right of=A4]{$S_3$};
  \node (D54) [above of=D6] {$\Delta(54)/\Z_3$};
  \node (S36) [above of=D54] {$\underline{\Sigma(36)}$};
  \draw[->] (E) to node {} (Z2);
  \draw[->] (E) to node {} (Z3);
  \draw[->] (Z2) to node {} (Z4);
  \draw[->] (Z2) to node {} (Z2Z2);
  \draw[->] (Z2) to node {} (D6);
  \draw[->] (Z3) to node {} (D6);
  \draw[->] (Z3) to node {} (A4);
  \draw[->] (Z4) to node {} (D8);
  \draw[->] (Z2Z2) to node {} (D8);
  \draw[->] (Z2Z2) to node {} (A4);
  \draw[->] (A4) to node {} (S4);
  \draw[->] (D8) to node {} (S4);
  \draw[->] (D6) to node {} (S4);
  \draw[->] (D6) to node {} (D54);
  \draw[->] (D54) to node {} (S36);
  \draw[->] (Z4) to node {} (S36);
\end{tikzpicture}
\caption{Tree of finite realizable groups of Higgs-family transformations in 3HDM.
Groups leading to automatic explicit CP-conservation are 
underlined. An arrow from $A$ to $B$ indicates that $A\subset B$.
}
   \label{fig-tree}
\end{figure}
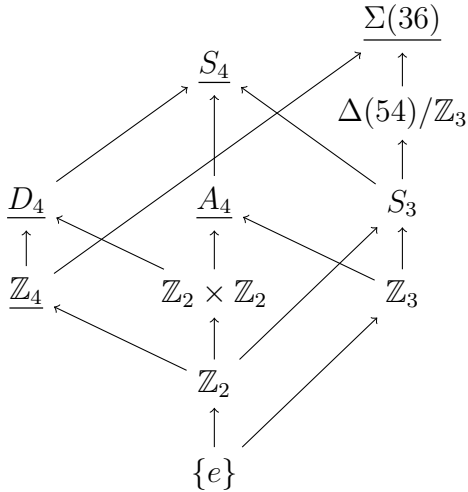

In 3HDM, we can have larger discrete symmetries, including several non-abelian groups.
The complete classification of realizable finite symmetry groups in 3HDM 
was achieved only very recently, \cite{classification3HDM}. 
If we focus on unitary transformations only, then there are ten realizable groups:
\bea
&&\Z_2, \quad \Z_3,\quad \Z_4,\quad \Z_2\times\Z_2,\quad D_3\simeq S_3,\quad D_4, \quad T\simeq A_4,\quad O\simeq S_4\,, \nonumber\\
&& (\Z_3\times\Z_3)\rtimes \Z_2 \simeq \Delta(54)/\Z_3,\quad (\Z_3\times\Z_3)\rtimes \Z_4\simeq \Sigma(36)\,.
\label{groups1}
\eea
This list is complete: trying to impose any other finite symmetry group of Higgs-family transformations
leads to a potential with continuous symmetry.
Fig.~\ref{fig-tree} should help visualize relations among different groups from this list.

The same work \cite{classification3HDM} also investigated
the relation between GCP symmetries and the Higgs-family symmetry groups.
It was found that, unlike in 2HDM, certain finite groups
do not automatically lead to the explicit $CP$-conservation. 
These are $\Z_2$, $\Z_3$, $\Z_2\times\Z_2$, $S_3$, and $\Delta(54)/\Z_3$. 
3HDM models based on them can be either explicitly $CP$-conserving or $CP$-violating.
However, the presence of $\Z_4$ or $A_4$ symmetry unavoidably leads to the explicit $CP$-conservation.

In what concerns symmetry breaking, we first remark that minimization of generic 
multicomponent
scalar potentials is a challenging task. In 3HDM, although the minimization problem 
can be formulated algebraically \cite{Nishi2007,Maniatis2014} or geometrically \cite{IvanovNHDM-II}, 
no known method is capable of solving it analytically in the general case.
For the case of symmetric 3HDMs, only specific examples tractable with straightforward algebra have been considered in literature. 
Beyond this case-by-case treatment, 
no systematic and exhaustive study for all of these groups in 3HDM exists.
It is the purpose of the present paper to fill this gap.

\subsection{Group-theoretic properties of breaking discrete symmetries}
\label{sec:break.large}

Let us now mention some group-theoretical observations 
which accompany breaking of a discrete symmetry group $G$ to its subgroup $G_v$
preserved by the vacuum.

First, electroweak symmetry breaking (EWSB) can lead to three outcomes:
either (i) the group is fully conserved by the vev alignment, $G_v = G$,
or (ii) the group is broken to a proper subgroup: $\{e\} \subset G_v \subset G$,
or (iii) the symmetry is broken completely, $G_v = \{e\}$.
Our goal in this paper is to establish all possible options for $G_v$ for each $G$ allowed in 3HDM.
Put simply, we want to establish the minimal and maximal amount of symmetry breaking for each $G$.
We will find that, for sufficiently large groups $G$, only option (ii) is available.

Second, there exists a relation between the number of degenerate global minima 
and the orders of the groups $G$ and $G_v$.
Suppose the discrete symmetry group $G$ is broken to $G_v \subset G$.
Let us denote the vev alignments corresponding to the global minima of the potential 
by $x_a \equiv (\langle\phi_1^0\rangle,\,\dots,\, \langle\phi_N^0\rangle)_a$, where $a$ runs from 1 to the total number 
of degenerate global minima $n$. The set of all $x_a$ is denoted as $X$,
on which the group $G$ acts by permutations. 
We can take any $x_a$ and observe that transformation $g \in G$ 
either keeps $x_a$ invariant, if $g \in G_v$, or sends it to another vev alignment $x_b$, if $g \not \in G_v$. 
In group-theoretic terms, $G_v$, being the subgroup of $G$ which keeps the chosen $x_a$ invariant, 
is known as the stabilizer (or little-group) of $x_a$.

If we start with a given alignment $x_a$ and
act with all $g \in G$, we obtain a $G$-orbit of length $\ell$ to which $x_a$ belongs.
The entire set $X$ is then partitioned into one or several disjoint orbits.
All $x_a$'s lying within any single orbit share the property that their stabilizers are isomorphic:
$G_v(x_a) \simeq G_v(x_b)$, although they can, in general, be different subgroups of $G$.
Stabilizers of vev alignments belonging to distinct orbits can be non-isomorphic.
Finally, within any single orbit, the following relation holds:
\be
\ell = |G|/|G_v|\,,
\ee
which is known in basic group theory as the orbit-stabilizer theorem.
If it happens that the set of global minima $X$ is covered by a single orbit, then $|G|/|G_v|$
is equal to the number of minima $n$. If $X$ contains more than one orbit, then $n = \sum_i \ell_i$.

\section{Symmetry breaking in 3HDM: groups without triplet irreps}\label{section-main1}


In this Section, we discuss the symmetry breaking features for discrete symmetry 
groups with Higgses in the singlet or doublet irreducible representations (irreps). 
To present them in a uniform fashion, we first introduce the notation for the 
group generators.
It is convenient to present each group in the basis where one of its abelian subgroups 
corresponds to rephasing transformations. In this basis, we will use the following generators:
\bea
\mbox{order 2:} &&
\sigma_{12} = \mathrm{diag}(-1,\,-1,\,1)\,,\quad
\sigma_{23} = \mathrm{diag}(1,\,-1,\,-1)\,,\quad 
c = -\mtrx{1&0&0\\ 0&0&1\\ 0&1&0}\,,\label{order2}\\
\mbox{order 3:} &&
a_3 = \mathrm{diag(1,\,\omega,\,\omega^2)}\quad\mbox{with}\quad\omega = \exp\left({2\pi i \over 3}\right)\,, \quad
b = \mtrx{0&1&0\\ 0&0&1\\ 1&0&0}\,,\label{order3}\\
\mbox{order 4:} &&
a_4 = \mathrm{diag}(1,\,i,\,-i)\,,\quad
d = {i \over\sqrt{3}} \left(\begin{array}{ccc} 1 & 1 & 1 \\ 1 & \omega^2 & \omega \\ 1 & \omega & \omega^2 \end{array}\right)\,.
\label{order4}
\eea
One can see that all abelian symmetries, 
$\Z_2,\Z_3,\Z_4,\Z_2\times\Z_2$ are represented by diagonal matrices.
Note also that $\sigma_{13}=\sigma_{12}\sigma_{23}$,
$a_4^2 = \sigma_{23}$, and $d^2 = c$.
The usual $CP$-transformation $\phi_i \mapsto \phi_i^*$ will be denoted simply by $CP$.
GCP transformations will be defined as $CP$ acting first, and then followed by a 
unitary transformation.

\tikzset{node distance=1.2cm, auto}
\begin{figure}[!htb]
   \centering
\begin{tikzpicture}
  \node (E) {$\{e\}$};
    \node (Z2) [above of=E] {$\langle\sigma_{23}\rangle$};
    \node (Z2Z2) [above of=Z2] 
{$\langle\sigma_{23},\sigma_{12}\rangle$};
    \node (Z3) [node distance=3cm, right of=Z2Z2] {$\langle a_3\rangle$};
    \node (Z4) [node distance=3cm, left of=Z2Z2] {$\langle a_4\rangle$};
    \node (D8) [above of=Z4] {$\langle a_4,c\rangle$};
    \node (A4) [above of=Z2Z2] {$\langle\sigma_{12},\sigma_{23},b\rangle $};
    \node (S4) [node distance = 1.8cm, above of=A4] {$\langle\sigma_{12},\sigma_{23},b,c\rangle $};
    \node (D6)[node distance=3cm, right of=A4]{$\langle a_3,c\rangle$};
    \node (D54) [above of=D6] {$\langle a_3,c,b\rangle$};
    \node (S36) [above of=D54] {$\langle a_3,c,b,d\rangle$};
  \draw[->] (E) to node {} (Z2);
  \draw[->] (E) to node {} (Z3);
  \draw[->] (Z2) to node {} (Z4);
  \draw[->] (Z2) to node {} (Z2Z2);
  \draw[->] (Z2) to node {} (D6);
  \draw[->] (Z3) to node {} (D6);
  \draw[->] (Z3) to node {} (A4);
  \draw[->] (Z4) to node {} (D8);
  \draw[->] (Z2Z2) to node {} (D8);
  \draw[->] (Z2Z2) to node {} (A4);
  \draw[->] (A4) to node {} (S4);
  \draw[->] (D8) to node {} (S4);
  \draw[->] (D6) to node {} (S4);
  \draw[->] (D6) to node {} (D54);
  \draw[->] (D54) to node {} (S36);
  \draw[->] (Z4) to node {} (S36);
\end{tikzpicture}
\caption{
Generating sets for the groups from Fig.\,\ref{fig-tree} in terms of generators 
given in Eqs.\,\eqref{order2}, \eqref{order3} and 
\eqref{order4}. 
In a slight abuse of notation, we show not the minimal generating sets
but the sets that should help visualize the construction of each group. 
Note also that these sets are not unique.}

   \label{fig-tree:gen}
\end{figure}
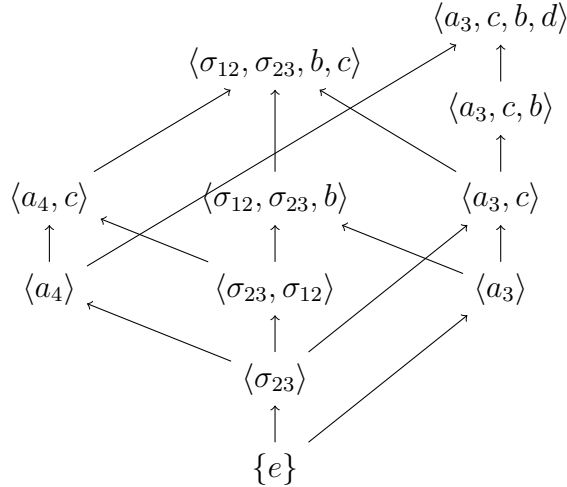

\subsection{Abelian Higgs-family groups}\label{subsection-abelian}

We begin with models based on abelian Higgs-family symmetry groups, 
both with and without explicit $CP$-violation:
\bea
\mbox{$CP$-violating:} && \Z_2, \quad \Z_3, \quad \Z_2 \times \Z_2\,,\nonumber\\
\mbox{$CP$-conserving:} && \Z_2^*, \quad \Z_2\times \Z_2^*, \quad \Z_4^*, \quad \Z_3 \rtimes \Z_2^*, \quad  \Z_4 \rtimes \Z_2^*, \quad
\Z_2 \times \Z_2\times \Z_2^*\,. \label{abeliangroups}
\eea
Here, an asterisk indicates that the generator of the corresponding group is a GCP transformation.
In the groups above, $\Z_2^*$ is generated by the usual CP transformation.
Usually $\Z_2^*$ will denote the presence of usual CP symmetry if otherwise not 
stated.
Such a distinction is relevant as the inclusion of different GCP symmetries leads 
to different groups. For example, the GCP transformation $c\cdot CP$ commutes 
with $\Z_3$ and $\Z_4$ so that we could define direct products $\Z_3\times\Z_2^*$ 
and $\Z_4\times\Z_2^*$, respectively.
These two groups, however, are shown to be non-realizable in 
3HDM \cite{abelianNHDM},
that is, they automatically leads to a larger Higgs family symmetry group.

The Higgs potentials can be generically written as $V = V_0 + V_{ph}$,
where $V_0$ is invariant under any phase rotation of individual doublets,
\be
V_0 =  - \sum_{1\le i \le 3} m_i^2(\phi_i^\dagger \phi_i) + \sum_{1 \le i\le j \le 3} \lambda_{ij} (\phi_i^\dagger \phi_i)(\phi_j^\dagger \phi_j)
+ \sum_{1 \le i < j \le 3} \lambda'_{ij} (\phi_i^\dagger \phi_j)(\phi_j^\dagger \phi_i)\label{V0}
\ee
with generic free parameters, while the phase-sensitive part $V_{ph}$ depends on the symmetry group $G$.
When evaluated at a generic neutral point with $\lr{\phi_i^0} = 
v_i e^{i\xi_i}/\sqrt{2}$, $V_0$ can be expressed
in terms of real non-negative quantities $r_i = v_i^2$ and with an obvious redefinition of the free parameters:
\be
V_0 = - M_i r_i + {1\over 2}\Lambda_{ij} r_i r_j = {1\over 2}\Lambda_{ij} (r_i - a_i) (r_j - a_j) + \mathrm{const}\,,\quad a_i = (\Lambda^{-1})_{ij} M_j.
\ee
The latter expression makes it clear that by appropriately choosing free parameters and adjusting the vector $a_i$, 
one can obtain $V_0$ whose global minimum has any values of $v_1$, $v_2$, $v_3$.
In particular, vev alignments such as $(v,\, 0,\, 0)$ and $(v_1,\, v_2,\, 0)$ are always possible.

The phase-sensitive part $V_{ph}$ selects out specific phases $\xi_i$, and it can also shift the values of $v_i$ just obtained.
However, in all cases of symmetry groups $G$, it is possible to identify a doublet, say $\phi_1$, such that
no quartic term in $V_{ph}$ contains three $\phi_1$'s and no quadratic term contains 
a single $\phi_1$.
As a result, the vev alignment of type $(v,\, 0,\, 0)$, found by minimizing $V_0$ 
only, remains stable upon inclusion of $V_{ph}$.
If we work with a symmetry group $G$ under which $\phi_1$ is a singlet, we conclude 
that such a vev alignment conserves the entire $G$.
For the alignment $(v_1,v_2,0)$ this argument does not work, and one needs to 
resort to other methods.

\subsubsection{Small groups} 

The $CP$-violating abelian groups, $\Z_2$, $\Z_3$, $\Z_2\times \Z_2$
(the latter is known as Weinberg's 3HDM \cite{weinberg3HDM}), can be broken, fully or partially,
by choosing appropriate $v_i$, which is made possible by the large number of free parameters.
As for $CP$-conserving models, breaking of $\Z_2^*$ 3HDM (no Higgs-family symmetry, just explicit $CP$-conservation)
and $\Z_2\times \Z_2^*$ 3HDM can be verified by straightforward calculations and follows from the fact that these models
possess phase-sensitive terms both in quadratic and in quartic parts of the potential.

The group $\Z_4^*$, despite being small, is somewhat special because it features a GCP transformation
whose square is not unity but a sign flip, for example, $\sigma_{23}$.
The generator of this transformation can be written as $q\cdot CP$, where
\be
q = \mtrx{1&0&0\\ 0&0&1\\ 0&-1&0}\,.\label{q}
\ee
Since $q$ is non-diagonal, it places certain restrictions on the parameters of $V_0$,
but the phase-sensitive part of the potential $V_{ph}$ is still sufficiently rich \cite{abelianNHDM}.
The overall number of free parameters remains large, two of them being complex with arbitrary phases,
which allows for any pattern of symmetry breaking of this group: 
the symmetry can be conserved, it can break to $\Z_2$, or it can break completely.

\subsubsection{Group $\Z_2 \times \Z_2 \times \Z_2^*$}

This is the $CP$-conserving version of the Weinberg's 3HDM \cite{weinberg3HDM},
which was first investigated in detail by Branco \cite{branco1980}.
Still, we present here a detailed discussion of its breaking patterns to facilitate exposition of
more complicated symmetries.

This group is based on sign flips and $CP$ and can be implemented via
the following phase-sensitive part of the potential:
\be
V_{ph} = \lambda_1 (\phi_2^\dagger \phi_3)^2 + \lambda_2 (\phi_3^\dagger \phi_1)^2 + \lambda_3 (\phi_1^\dagger \phi_2)^2 + h.c.
\label{VphZ2Z2Z2}
\ee
with non-zero real $\lambda_i$. Writing the generic vev alignment and differentiating it with respect to the phases $\xi_i$
yields:
\be
\lambda_1 r_2 r_3 \sin(2\xi_2-2\xi_3) = \lambda_2 r_3 r_1 \sin(2\xi_3-2\xi_1) = \lambda_3 r_1 r_2 \sin(2\xi_1-2\xi_2)\,.\label{sineZ2Z2Z2}
\ee
Now, several situations are possible. 
\begin{itemize}
\item
If two vevs are zero, the phase condition is irrelevant, and we get the alignment of type $(v,\,0,\,0)$ which conserves the full group $G$.
\item
If one of vevs is zero, $r_1 = 0$, then we have a freedom of shifting $\xi_2+\xi_3$. We can set it to zero,
and then obtain $\xi_2 = -\xi_3 = \pi k/4$. The values of $k$ corresponding to a minimum depend on the sign of $\lambda_1$:
if $\lambda_1 > 0$, then $k$ must be odd, if $\lambda_1 < 0$, then $k$ must be even.
In each of these two cases, we obtain only two distinct vev alignments:
\be
\mbox{if }\lambda_1 > 0: \quad (0,\, v_2 e^{i\pi/4},\, \pm v_3 e^{-i\pi/4})\,;\qquad 
\mbox{if }\lambda_1 < 0: \quad (0,\, v_2 ,\, \pm v_3 )\,.
\ee
Each minimum is symmetric under $\sigma_{23}$ and a GCP symmetry,
and the generator corresponding to the broken symmetry, $\sigma_{12}$, links the two minima. 
The residual symmetry group is $G_v = \Z_2 \times\Z_2^*$;
the two minima form a single orbit of length $\ell = |G|/|G_v| = 2$.
\item
Another possibility is that all $v_i \not = 0$ but the sines are zero.
Then (\ref{sineZ2Z2Z2}) can still be satisfied when all $\xi_i = \pi k_i/2$.
Again, the specific choices for $k_i$ depend on the signs of $\lambda_i$.
Still, in each case, the residual symmetry group is just $G_v = \Z_2^*$ (either $CP$ or a GCP transformation), 
and we have four minima sitting on a single orbit.
\item
Finally, we can search for non-zero solutions of (\ref{sineZ2Z2Z2}). 
These equations can be viewed as the law of sines written for the triangle
with sides $L_i = r_i/|\lambda_i|$ and angles $\alpha_1 = \pi- 
(2\xi_2-2\xi_3)$, $\alpha_2 = \pi- (2\xi_3-2\xi_1)$, $\alpha_3 = 
\pi - (2\xi_1-2\xi_2)$, for positive $\lambda_i$; for negative $\lambda_i$, one 
should subtract $\pi$ from the expression of $\alpha_i$.
Notice also that the angles $\alpha_i$ should be taken modulo $2\pi$.
If the sides satisfy the obvious inequalities, the angles
$\alpha_i$ are determined uniquely,
and one then finds the phases $\xi_i$. Since these phases are {\em generic},
they completely break the symmetry group: $G_v = \{e\}$.
As a result, we obtain eight minima lying on a single orbit of length $\ell = |G|/|G_v| = 8$.
The possibility of spontaneous $CP$-violation in this model was already 
mentioned in the original Weinberg's paper \cite{weinberg3HDM}.
\end{itemize}
In short, we find that all symmetry breaking patterns are possible for $G = \Z_2\times\Z_2\times\Z_2^*$.

\subsubsection{Group $\Z_3 \rtimes \Z_2^*$}

The $Z_3$-symmetric 3HDM is based on generic $V_0$ and on the following $V_{ph}$:
\be
V_{ph} = \lambda_1 (\phi_2^\dagger \phi_1)(\phi_3^\dagger \phi_1) + 
\lambda_2 (\phi_3^\dagger \phi_2)(\phi_1^\dagger \phi_2) 
+ \lambda_3 (\phi_1^\dagger \phi_3)(\phi_2^\dagger \phi_3) + h.c.\label{VphZ3}
\ee
The $\Z_3$ symmetry group is generated by $a_3$.
When all $\lambda_i$ are real, the model is also $CP$-conserving, and the full symmetry group is extended to 
$\Z_3 \rtimes \Z_2^* \simeq S_3$.
Whether $\Z_3$ is broken at the minimum or not depends on non-zero values $v_i$.
Either variant is possible with a suitable $V_0$.
However it remains to be studied what phases $\xi_i$ the vevs can acquire, and whether the vacuum is invariant
under any GCP transformation.

Evaluating the potential at a generic vev alignment and differentiating with respect to phases $\xi_i$,
we obtain two equalities 
\be
\lambda_1 v_1^2 v_2 v_3\sin(2\xi_1-\xi_2-\xi_3) = 
\lambda_2 v_2^2 v_1 v_3\sin(2\xi_2-\xi_1-\xi_3) = 
\lambda_3 v_3^2 v_1 v_2\sin(2\xi_3-\xi_1-\xi_2)\,.\label{sineZ3Z2}
\ee
which resemble (\ref{sineZ2Z2Z2}) but have different phase arrangements.
If all three $v_i$ are non-zero, we can again search for zero and non-zero solutions
of (\ref{sineZ3Z2}). 
To get non-zero solutions, we interpret (\ref{sineZ3Z2})
as the law of sines for a triangle with side $L_i = (|\lambda_i| v_i)^{-1}$ and
angles $\alpha_1 = \pi- 2\xi_1 + \xi_2 + \xi_3$ (for positive $\lambda_1$), etc. 
Once again, the values of $L_i$ fix $\alpha_i$, which in turn give the values of 
$\xi_i$.
These values are determined uniquely, up to group transformations.
They are not rigid, as they continuously change upon variation of free parameters,
and they break the $CP$-symmetry. Thus, in this case we obtain six minima
differing by phases with no residual symmetry. 
They lie on a single orbit of length $\ell = |G|/|G_v| = 6$.

Zero solutions to (\ref{sineZ3Z2}) arise when $v_i$'s are such that the triangle with given sides
does not exist. Then, the phases of vevs are aligned, up to the group transformation,
and the minimum conserves the $CP$ symmetry.

In short, the symmetry group $\Z_3 \rtimes \Z_2^*$ can be either conserved,
or broken, either to $\Z_2^*$ or completely. Note that there is no way to break only 
$CP$ but keep the $\Z_3$ symmetry.

\subsubsection{Group $\Z_4 \rtimes \Z_2^*$}

The $Z_4$-symmetric 3HDM is based on generic $V_0$ and on the following $V_{ph}$:
\be
V_{ph} = \lambda_1 (\phi_2^\dagger \phi_1)(\phi_3^\dagger \phi_1) + 
\lambda_2 (\phi_3^\dagger \phi_2)^2 + h.c.\label{VphZ4}
\ee
The $\Z_4$ symmetry group is generated by $a_4$.
Since we have only two free parameters here, we can rephase the doublets in such a way that
$\lambda_1$ and $\lambda_2$ become real.
Thus, this model is also $CP$-conserving, and the full symmetry group is extended to 
$\Z_4 \rtimes \Z_2^* \simeq D_4$.

Also, since we now have two, not three free parameters in $V_{ph}$, differentiating in $\xi_i$
gives
\be
v_1^2 v_2 v_3 \sin(\xi_2+\xi_3) = 0\,,\quad 
v_2^2 v_3^2 \sin(2\xi_2 - 2\xi_3) = 0\,.\label{sineZ4}
\ee
In these equations, we already set $\xi_1 = 0$. 
Suppose first that $v_1 =0$ and $v_2, v_3 \not = 0$. We then obtain exactly the same situation 
as was considered as the second option for the group 
$\Z_2\times \Z_2 \times \Z_2^*$.
We get a vev alignment with phases which are certain multiples of $\pi/4$
depending on the sign of $\lambda_2$.
Each minimum will be symmetric under $\sigma_{23}$ and a GCP transformation,
which generate $G_v = \Z_2 \times \Z_2^*$. 
There are two minima linked by $a_4$, which sit on a single orbit of length two:
$(0,v_2,v_3),(0,v_2,-v_3)$.

If all $v_i \not = 0$, the situation does not change much.
We still get $\xi_2 = \pi k/4$, $\xi_3 = -\xi_2 + \pi p$.
The signs of $\lambda_1$ and $\lambda_2$ fix whether integers $k$ and $p$
are even or odd, and in each case we still find that the vev alignment is symmetric under a GCP
transformation. However, the non-zero value of $v_1$ makes difference:
the $\sigma_{23}$ symmetry is now absent and it links instead distinct pairs of minima.
Thus, we get $G_v = \Z_2^*$ and a single orbit with four minima:
$(v_1,\pm v_2e^{\pm i\xi_2},\mp v_3e^{\mp i\xi_2})$.

The last possibility is when $v_1, v_2 \not = 0$, while $v_3 = 0$.
It leads to the alignment $(v_1,\, v_2 e^{i\xi_2},\,0)$ with arbitrary $\xi_2$.
However the structure of the hessian makes it a saddle point rather than a minimum,
due to $\partial^2 V/\partial \xi_2^2 = 0$ but $\partial^2 V/\partial v_3 \partial \xi_2 \not = 0$.

In short, we find the following breaking patterns for the group $G = \Z_4 \rtimes \Z_2^*$:
$G$ can be conserved, or it can be broken, either to $\Z_2\times \Z_2^*$ or to $\Z_2^*$.
Similarly to the $\Z_3 \rtimes \Z_2^*$ case, there is no way to break the $CP$-symmetry
keeping the Higgs-family symmetry intact. In contrast to that case,
it is now impossible to break $\Z_4 \rtimes \Z_2^*$ {\em completely},
and a GCP symmetry is always preserved by the vacuum. This leads us to the following conclusion:
$\Z_4$ Higgs-family symmetry protects the 3HDM scalar sector
from {\em any form of $CP$-violation}, either explicit or spontaneous.

\subsection{Groups $S_3$ and $D_4$}

List (\ref{groups1}) contains two non-abelian Higgs-family groups, $S_3$ and $D_4$,
whose irreducible representations are only singlets and doublets.
Note that $S_3$-symmetric 3HDM can be $CP$-violating and $CP$-conserving; in the latter
case the full symmetry group of the model is $S_3\times \Z_2^*$. 
In contrast, $D_4$-symmetric 3HDM can only be $CP$-conserving, with symmetry group 
$D_4 \times \Z_2^*$. Note that semidirect products of groups became direct products with 
the aid of a $\Z_2^*$ generator $c \cdot CP$: for example, $(c \cdot CP)^{-1}a_3 (c \cdot CP) = a_3$.

Let us start with observations applicable to all these groups.
Since they contain a singlet, for example $\phi_1$, we can 
repeat the arguments of Section~\ref{subsection-abelian}: 
by adjusting the coefficients of the potential, one can guarantee 
that the vev alignment $(v,\, 0,\, 0)$ becomes the global minimum and conserves the entire group $G$.
Thus, for each of these groups, the minimal amount of symmetry breaking is no breaking at all.

These symmetry groups can be viewed as extension of models considered above 
extended by the additional generator $c$.
The irreducible representations correspond to $\bs{1}_0+\bs{2}$ for both $S_3$ and 
$D_4$ ($\bs{1}_0$ is the trivial invariant) but all representations 
$\bs{1}_i+\bs{2}$ with nontrivial singlet $\bs{1}_i$ are equivalent by basis 
change when we can factor the global $U(1)$.
Invariance under $c$ imposes restrictions both on $V_0$ and $V_{ph}$.
The phase-invariant part of the potential can be written as
\bea
V_0 &=& - m_1^2 (\phi_1^\dagger \phi_1) - m_2^2 \left(\phi_2^\dagger \phi_2 + \phi_3^\dagger \phi_3\right)
+ {1\over 2}\lambda_1 (\phi_1^\dagger \phi_1)^2 +  {1\over 2}\lambda_2 \left[(\phi_2^\dagger \phi_2)^2 + (\phi_3^\dagger \phi_3)^2 \right]
\nonumber\\
&& + \lambda_3  (\phi_1^\dagger \phi_1) \left(\phi_2^\dagger \phi_2 + \phi_3^\dagger \phi_3 \right) + 
\lambda_4 (\phi_2^\dagger \phi_2) (\phi_3^\dagger \phi_3)  + \lambda'_3 (z_{12} + z_{13}) + \lambda'_4 z_{23} \,,
\label{V0new}
\eea
where $z_{ij} \equiv (\phi_i^\dagger \phi_i)(\phi_j^\dagger \phi_j) - (\phi_i^\dagger \phi_j)(\phi_j^\dagger \phi_i)$,
and all three $z_{ij} \ge 0$ are algebraically independent \cite{IvanovNishiNHDM}.
The $V_{ph}$ part of the potential must also incorporate the $\phi_2 \leftrightarrow \phi_3$ symmetry.
In the case of $D_4$, $V_{ph}$ is the same as in (\ref{VphZ4}), while for the $S_3$ we have 
a simplified version of (\ref{VphZ3}):
\be
V_{ph} = {1 \over 2}\lambda_5 (\phi_2^\dagger \phi_1)(\phi_3^\dagger \phi_1) + 
{1 \over 2}\lambda_6 \left[ (\phi_3^\dagger \phi_2)(\phi_1^\dagger \phi_2) 
+ (\phi_1^\dagger \phi_3)(\phi_2^\dagger \phi_3)\right] + h.c.\label{VphS3}
\ee
We get the $CP$-conserving version of the $S_3$ model when $\lambda_5$ and $\lambda_6$ are real.

Next, let us check whether these groups can be broken completely by minimization of the potential.
In an attempt to do so, we need to break, among other, two order-2 symmetries: $\phi_2 \leftrightarrow \phi_3$ and $CP$.
These symmetries interact via $V_{ph}$, and it is not clear {\em a priori} that there exists a minimum
which breaks both of them. In Appendix~\ref{appendix-D4}, using a geometric reinterpretation of the extremization problem,
we show that either we get trivial phases which protect $CP$, or we get $v_2= v_3$ and correlated 
(equal or opposite) phases, which results in a $\Z_2$ or GCP residual symmetry. 
Thus, these alignments break the groups $D_4 \times \Z_2^*$ and $S_3 \times \Z_2^*$ to $G_v = \Z_2$ or $\Z_2^*$.
In each case, we have 8 or 6 minima lying on a single orbit,
and the complete breaking of these symmetry groups is not feasible in 3HDM.

Since the $CP$-conserving $S_3$ 3HDM allows for a minimum with 
$v_2 \not = v_3$, such a solution is also possible for its explicitly $CP$-violating version.
In this case, the phases are irrelevant, and the entire symmetry group $S_3$ is broken.
A partial breaking to $\Z_2$ by alignment $v_2 = v_3$, $\xi_2 = \xi_3$, 
or to $\Z_3$ by alignment $(0,\, v,\, 0)$ are also possible;
for cross-check, we verified these possibilities with numerical examples.

\section{Groups with triplet irreps}\label{section-main2}

\subsection{$A_4$ and $S_4$}

The group $A_4$ has received a lot of attention in the bSM literature \cite{3HDM-A4},
in part because it is the smallest finite group possessing a three-dimensional irrep.
Following \cite{minimization2012}, we write the $A_4$-symmetric 3HDM potential in the following way:
\bea
V&=&-\frac{M_0}{\sqrt{3}}\left(\phi_1^{\dagger}\phi_1+\phi_2^{\dagger}\phi_2+\phi_3^{\dagger}\phi_3\right)+\frac{\Lambda_0}{3}\left(\phi_1^{\dagger}\phi_1+\phi_2^{\dagger}\phi_2+\phi_3^{\dagger}\phi_3\right)^2\nonumber\\ 
&&+\frac{\Lambda_3}{3}\left[(\phi_1^{\dagger}\phi_1)^2+(\phi_2^{\dagger}\phi_2)^2+(\phi_3^{\dagger}\phi_3)^2-(\phi_1^{\dagger}\phi_1)(\phi_2^{\dagger}\phi_2)-(\phi_2^{\dagger}\phi_2)(\phi_3^{\dagger}\phi_3)-(\phi_3^{\dagger}\phi_3)(\phi_1^{\dagger}\phi_1)\right]\nonumber\\
&&+\Lambda_1\left[(\Re\phi_1^{\dagger}\phi_2)^2+(\Re\phi_2^{\dagger}\phi_3)^2+(\Re\phi_3^{\dagger}\phi_1)^2\right]\nonumber\\
&&+\Lambda_2\left[(\Im\phi_1^{\dagger}\phi_2)^2+(\Im\phi_2^{\dagger}\phi_3)^2+(\Im\phi_3^{\dagger}\phi_1)^2\right]\nonumber \\
&&+\Lambda_4\left[(\Re\phi_1^{\dagger}\phi_2)(\Im\phi_1^{\dagger}\phi_2)+(\Re\phi_2^{\dagger}\phi_3)(\Im\phi_2^{\dagger}\phi_3)+
(\Re\phi_3^{\dagger}\phi_1)(\Im\phi_3^{\dagger}\phi_1)\right]\,,
\label{Tetrahedralgeneral}
\eea
with generic real parameters $M_0$ and $\Lambda_i$.
It is symmetric under the group $A_4$ of Higgs-family transformations 
generated by independent sign flips of individual doublets $\sigma_{12}, \sigma_{23}$ and
by the cyclic permutation $b$. 
It is also automatically symmetric under GCP transformation generated, for example, by $c\cdot CP$.
The full symmetry group of this potential is therefore $G = A_4 \rtimes \Z_2^*$ of order 24.
For generic values of the parameters, this potential has no other Higgs-family or GCP symmetries.

Minimization of this potential was investigated in full detail in \cite{minimization2012},
with the aid of a geometrical method. The global minima can have the following four vev alignments $(v_1,\, v_2,\, v_3)$:
\be
A= (1,0,0)\,,\quad B= (1,1,1)\,,\quad C = (\pm 1,\, \omega,\, \omega^2)\,,\quad D = (0,\, 1,\, e^{i\alpha})\,,
\label{A4minima}
\ee
where the overall vev scale is factored out and
\be
\sin 2\alpha = -{\Lambda_4 \over \sqrt{(\Lambda_1 - \Lambda_2)^2 + \Lambda_4^2}}\,,\quad 
\cos 2\alpha = -{\Lambda_1 - \Lambda_2 \over \sqrt{(\Lambda_1 - \Lambda_2)^2 + \Lambda_4^2}}\,. \quad
\alpha \not = {\pi \over 3} k\,.
\ee
The values of $\alpha$ equal to multiples of $\pi/3$ are excluded because in these cases point D
leads to an unwanted massless scalar.
Among the four cases, the first three are rigid in the sense that the global minimum is insensitive 
to moderate variation of the free parameters, while the last one is flexible. 
To avoid misunderstanding, in all cases we imply not only the specific vev alignment written explicitly, 
but also other alignments which can be obtained from them by the broken symmetry generators.

Setting $\Lambda_4 = 0$ in Eq.~(\ref{Tetrahedralgeneral})
leads to the 3HDM symmetric under $G = S_4 \times \Z_2^*$ of order 48,
which is generated by $\sigma_{12}$, $\sigma_{23}$, $b$, $c$, and $CP$.
The global minima are almost identical to \eqref{A4minima}:
\be
A= (1,0,0)\,,\quad B= (1,1,1)\,,\quad C = (\pm 1,\, \omega,\, \omega^2)\,,\quad D = (0,\,1,\, i)\,.
\label{S4minima}
\ee
However since the symmetry groups $G$ are different in the two cases, 
the remaining symmetries $G_v$ at each vacuum might also differ.

Let us now take a closer look at the group-theoretic properties at each of these four minima.
\begin{itemize}
\item
Point $A$ is invariant under all sign flips, under $CP$ and $c$, but not under cyclic permutations generated by $b$.
For the $A_4$ 3HDM, the remaining group is $G_v =\langle \sigma_{12},\,\sigma_{23},\, c\cdot CP\rangle$ of order $|G_v|=8$.  
For the $S_4$ 3HDM, $G_v =\langle \sigma_{12},\,\sigma_{23},\, c,\, CP\rangle$ with $|G_v|=16$. 
In both cases there are $n = 3$ minima of type A all linked to each other by the broken generator $b$,
which form a single orbit: 
$(1,0,0),(0,1,0),(0,0,1)$ with $n = \ell = |G|/|G_v|$.
\item
Point $B$ is invariant under all permutations as well as $CP$, but not under sign flips.
The remaining group is $G_v =\langle b,\, c\cdot CP\rangle$ of order $|G_v|=6$ for the $A_4$ 3HDM and 
$G_v =\langle b,\, c,\, CP\rangle$ of order $|G_v|=12$ for the $S_4$ 3HDM. 
There are $n=4$ distinct degenerate minima $(1,\pm 1,\pm 1)$ 
which form a single orbit of length $\ell = |G|/|G_v| = 4$ 
and which are obtained from point $B$ by individual sign flips.
Note that vev alignments $(1,\,-1,\,-1)$ and $(-1,\,1,\,1)$ are considered the same because
they differ by an overall phase factor $-1$.
\item
Point $C$ includes two seemingly different sorts of vev alignments,
\bea
C_+ =  (1,\, \omega,\, \omega^2)\,,\quad C_- =  (-1,\, \omega,\, \omega^2)\,,\label{pointsCpm}
\eea
which are graphically shown on the complex plane in Fig.~\ref{fig-pointC}.
\begin{figure}[!htb]
   \centering
\includegraphics[width=10cm]{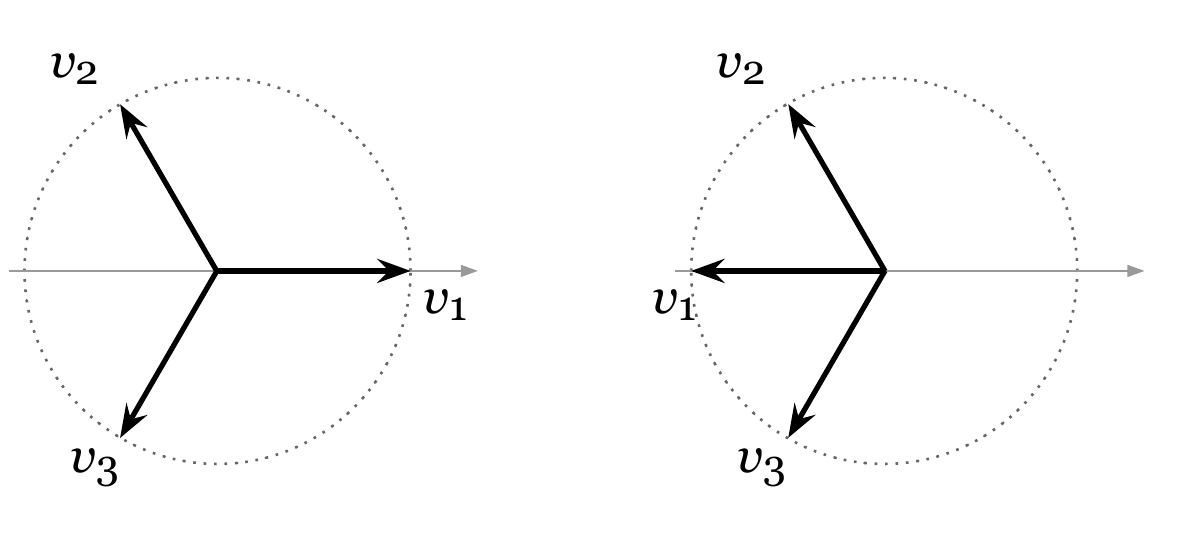}
\caption{Graphic representations of the vev alignments on the complex plane for points $C_+$ (left) and $C_-$ (right)}
   \label{fig-pointC}
\end{figure}
These pictures facilitate counting the number of minima and their residual symmetries.
Up to the overall phase factor, there are only two points of type $C_+$, the one given in Eq.~(\ref{pointsCpm})
and its conjugate, and six points of type $C_-$ which differ by permutations.
Despite an apparent difference, these two classes of points have isomorphic groups $G_v$.
$C_+$ is invariant under $G_v = \langle b, \, c\cdot CP\rangle$, 
while $C_-$ conserves $G_v = \langle \sigma_{13}\cdot b, \, c\cdot CP\rangle$.
In both cases, $G_v \simeq S_3$ of order 6, and its order-2 elements are GCP transformations.

A non-trivial fact is that these $G_v$'s apply to {\em both} $A_4$ and to $S_4$ symmetries. 
Under $G=S_4\times \Z_2^*$, these eight minima form a single orbit of length $\ell = |G|/|G_v| = 8$,
while under $G=A_4\rtimes \Z_2^*$, they split into {\em two disjoint orbits} of length $\ell = 4$, 
which add up to $n = 2\ell = 8$ minima:
$(1,\pm\omega,\pm \omega^2)$ and $(1,\pm\omega^2,\pm \omega)$.

\item
Point $D$ is invariant under one sign flip, 
and an exchange of two doublets accompanied either by the $CP$ transformation 
or by another sign flip (the latter possibility appears only in $S_4$).
The remaining group is $G_v =\langle \sigma_{23},\, c\cdot CP\rangle$ of order 4 for the $A_4$ 3HDM, 
and $G_v =\langle \sigma_{23},\, c\cdot \sigma_{13},\, c\cdot CP\rangle$ with $|G_v|=8$ for the $S_4$ 3HDM.
In both cases, we have six minima of type $D$ linked by the permutations and forming 
a single orbit [$(0,1,\pm i)$ and cyclic permutations] of length
$\ell = |G|/|G_v|=6$.
\end{itemize}
Note that for all four types of symmetry breaking, the vacuum is still invariant under a GCP symmetry,
which is a manifestation of the well-known fact that $A_4$ or $S_4$ symmetric 3HDM does not 
offer possibility neither for explicit nor spontaneous $CP$-violation \cite{3HDM-A4}.

\subsection{$\Delta(27)$ family}

The three remaining symmetry groups in 3HDM are
\be
G = (\Z_3\times \Z_3)\rtimes \Z_2 \simeq \Delta(54)/\Z_3\,,\quad (\Z_3\times \Z_3)\rtimes (\Z_2 \times \Z_2^*)\,,\quad
\Sigma(36)\rtimes \Z_2^*\,,\label{GDelta27}
\ee
of orders $|G|=18$, 36, and 72, respectively.
Here, as usual, an asterisk in $\Z_2$ denotes a GCP transformation.
We stress that, according to our discussion in Section~\ref{section-overview}, 
these are subgroups of $PSU(3) \equiv SU(3)/Z(SU(3))$, where $Z(SU(3)) = \Z_3$ is the center of $SU(3)$.
In order to keep the notation more familiar, 
we will call such models as ``$CP$-violating $\Delta(54)$'', 
``$CP$-conserving $\Delta(54)$'', and ``$\Sigma(36)$'' 3HDMs, --- that is, we will 
refer to their full preimages $\hat G \subset SU(3)$ rather than groups $G \subset 
PSU(3)$ themselves.
The accurate correspondence between them is listed in Table~\ref{table1}.
Collectively, we will call these groups the ``$\Delta(27)$ family'' because the corresponding 
$SU(3)$ preimages contain $\Delta(27)$ as a subgroup.
Note that the group $\Delta(27)$ itself is absent from the list (\ref{GDelta27}) because it is not realizable:
the scalar potential symmetric under $\Delta(27)$ is always symmetric under the larger group $\Delta(54)$.

\begin{table}[htb]
\centering
  \begin{tabular}{| c | c|c | c|c |}
\hline
label & $\hat G \subset SU(3)$ & $|\hat G|$ & $G \subset PSU(3)$ & $|G|$ \\   \hline\hline
$\Delta(27)$ group & $\Delta(27)$ & 27 & $\Z_3\times \Z_3$ & 9 \\
$CP$-violating $\Delta(54)$ 3HDM & $\Delta(54)$ & 54 & $(\Z_3\times \Z_3)\rtimes \Z_2$ & 18 \\
$CP$-conserving $\Delta(54)$ 3HDM & $\Delta(54)\rtimes \Z_2^*$ & 108 & $(\Z_3\times \Z_3)\rtimes (\Z_2 \times\Z_2^*)$ & 36 \\
$\Sigma(36)$ 3HDM & $\Sigma(36\phi)\rtimes\Z_2^*$ & 216 & $\Sigma(36)\rtimes \Z_2^*$ & 72 \\
\hline
  \end{tabular}
 \caption{Conventions for the symmetry groups from the $\Delta(27)$-family}\label{table1}
\end{table}

There exists a basis in which the group $\Delta(27)$ is generated by $a_3$ and $b$, and the $\Delta(27)$-symmetric
potential takes the form
\bea
V_1 & = &  - m^2 \left[\phi_1^\dagger \phi_1+ \phi_2^\dagger \phi_2+\phi_3^\dagger \phi_3\right]
+ \lambda_0 \left[\phi_1^\dagger \phi_1+ \phi_2^\dagger \phi_2+\phi_3^\dagger \phi_3\right]^2 \nonumber\\
&&+ {\lambda_1 \over 3} \left[(\phi_1^\dagger \phi_1)^2+ (\phi_2^\dagger \phi_2)^2+(\phi_3^\dagger \phi_3)^2
- (\phi_1^\dagger \phi_1)(\phi_2^\dagger \phi_2) - (\phi_2^\dagger \phi_2)(\phi_3^\dagger \phi_3)
- (\phi_3^\dagger \phi_3)(\phi_1^\dagger \phi_1)\right]\nonumber\\
&&+ \lambda_2 \left[|\phi_1^\dagger \phi_2|^2 + |\phi_2^\dagger \phi_3|^2 + |\phi_3^\dagger \phi_1|^2\right] \nonumber\\
&&+ \left(\lambda_3 \left[(\phi_1^\dagger \phi_2)(\phi_1^\dagger \phi_3) + (\phi_2^\dagger \phi_3)(\phi_2^\dagger \phi_1) + (\phi_3^\dagger 
\phi_1)(\phi_3^\dagger \phi_2)\right]
+ h.c.\right),\label{potential-Delta54}
\eea
with generic real $m^2$, $\lambda_0$, $\lambda_1$, $\lambda_2$ and a complex $\lambda_3$. 
It can be checked that this potential is also symmetric under $c$, so that the resulting symmetry group 
is $\Delta(54)= \langle a_3,\, b,\, c\rangle$. We refer to this model as $CP$-violating $\Delta(54)$ 3HDM.

If the free parameters in (\ref{potential-Delta54}) are generic, the potential is not symmetric under any other transformation, 
be it a Higgs-family or GCP transformation.
If $\lambda_3$ is real or can be made real by rephasing, which is possible when its phase is a multiple of $\pi/3$,
then the potential (\ref{potential-Delta54}) becomes invariant under the usual $CP$ transformation,
and the symmetry group is then promoted to $\Delta(54) \rtimes \Z_2^*$.
We refer to this model as $CP$-conserving $\Delta(54)$ 3HDM.

Finally, if $\lambda_3$, apart from just being real, satisfies $\lambda_3 = (\lambda_1 - \lambda_2)/2$,
the symmetry group of (\ref{potential-Delta54}) is enlarged to $\Sigma(36\phi) \rtimes \Z_2^*$ generated by $a_3$, $b$, $CP$
and the order-4 generator $d$ given in (\ref{order4}). 

We can find the global minima of the potentials in the $\Delta(27)$ family with the same geometric method
which was used in \cite{minimization2012} for the $A_4$ and $S_4$ groups.
Details of this analysis are given in Appendix~\ref{appendix-D27}; here we just summarize 
the results\footnote{Our results disagree with those obtained in \cite{Venus2014} where the same problem was addressed.}.
The possible global minima can have the following vev alignments:
\bea
\mbox{point $A$:} && (\omega,\,1,\,1)\,,\quad (1,\,\omega,\,1)\,,\quad(1,\,1,\,\omega)\,,\nonumber\\
\mbox{point $A'$:} && (\omega^2,\,1,\,1)\,,\quad (1,\,\omega^2,\,1)\,,\quad(1,\,1,\,\omega^2)\,,\nonumber\\
\mbox{point $B$:} && (1,\,0,\,0)\,,\quad (0,\,1,\,0)\,,\quad(0,\,0,\,1)\,.\nonumber\\
\mbox{point $C$:} && (1,\,1,\,1)\,,\quad (1,\,\omega,\,\omega^2)\,,\quad(1,\,\omega^2,\,\omega)\,.\label{AA'BC}
\eea
The three realizable groups from the $\Delta(27)$ family listed in Table~\ref{table1} can have the global minima only
at these values. These minima can be degenerate, and the higher the symmetry, the stronger is the degeneracy:
\bea
\mbox{$CP$-violating $\Delta(54)$ 3HDM}&&: \quad
A\,,\ \ A'\,,\quad B\,,\ \ C\,,\\
\mbox{$CP$-conserving $\Delta(54)$ 3HDM} &&: \quad
A + A'\,,\quad B\,,\ \ C\,,\\
\Sigma(36)~\text{3HDM}&&: \quad
A + A'\,,\quad B + C\,.
\eea
Here, symbol $+$ means that two points merge to a single point
in the corresponding lower-dimensional orbit space, and therefore the global minima at these points are denegerate.

Let us now investigate the residual symmetry groups $G_v$ for each of the three groups from
the $\Delta(27)$ family.
\begin{itemize}
\item
In the case of $CP$-violating $\Delta(54)$, it turns out that each possible vev alignment 
listed in (\ref{AA'BC}) is invariant, up to an overall phase factor, under a certain $G_v \simeq S_3$. 
For example, the first vev alignment of point $A$ in (\ref{AA'BC}) is invariant under $G_v=\lr{c, a_3 b^2}$, 
the first $A'$ point is invariant under $G_v=\lr{c,  a_3 b}$, the first $B$ point is invariant under $G_v=\lr{a_3, c}$,
and the first $C$ point is invariant under any permutation of doublets, $G_v=\nolinebreak\lr{b,c}$.
In all four cases, the minima form a single orbit of length $\ell = |G|/|G_v| = 3$. 

Thus, spontaneous breaking of this symmetry group looks remarkably simple in group-theoretic terms.
It correspond to removal of one of the four $\Z_3$ generators present in $G = (\Z_3\times\Z_3)\rtimes\Z_2$:
$a_3,\, a_3b,\, a_3b^2$, or $b$.
This broken $\Z_3$ symmetry links together the three global minima.

\item
In the case of $CP$-conserving $\Delta(54)$,
we have an additional symmetry generator: $CP$.
Simultaneously, the points $A$ and $A'$ fuse to a single point in the corresponding orbit space,
see details in Appendix~\ref{appendix-D27}. These points have the same $G_v$ as before
but now they form a single $G$-orbit of length $\ell = |G|/|G_v| = 6$.
In contrast, points $B$ and $C$ stay separately, each comprising one orbit of length $\ell = 3$.
In these cases, each vev alignment possesses an additional GCP symmetry. 
Thus, the residual symmetry group for points $B$ or $C$ has $|G_v|=12$,
so that $\ell = |G|/|G_v| = 3$.
In group-theoretic terms, spontaneous breaking of $G = (\Z_3\times\Z_3)\rtimes(\Z_2\times \Z_2^*)$
can proceed via removal of either one $\Z_3$ generator (points $B$ or $C$), 
or one $\Z_3$ and one $\Z_2$ generator (point $A+A'$).

Points $A + A'$ for this model serve as the prototypical case 
of the phenomenon of geometric $CP$-violation \cite{geometricCP}.
These minima spontaneously violate $CP$-symmetry of the model,
but the relative phase between vevs is fixed by geometric requirements 
and does not change as the parameters of the potential continuously change.
This behaviour is quite distinct from the $(0,\, 1,\, e^{i\alpha})$ alignment of the $A_4$ model,
where the phase $\alpha$ depends on the values of the parameters
but where there is not true spontaneous $CP$-violation.

We also note the curious fact that this relative phase is even {\em more robust}
than we could naively expect. Indeed, it was already present 
in the $CP$-violating $\Delta(54)$ model with an arbitrary value 
of the phase of $\lambda_3$. This fact goes in line with the general
observation made in \cite{minimization2012} 
that the orbit spaces of highly symmetric potentials tend to be very cuspy,
and their minimization leads to very rigid structures in vev alignments.

\item
In the case of $\Sigma(36)$ 3HDM, we have a new symmetry $d$ at our disposal.
When acting in the orbit space, it links together points $B$ and $C$, for example
\be
d\,(1,1,1)^{\mathrm{T}}=\sqrt{3}(1,0,0)^{\mathrm{T}}\,.
\ee
It also provides a second link between points $A$ and $A'$, complementing $CP$
and enlarging the residual symmetry group $G_v$. 
Thus, we are left with only two sets of degenerate minima, $A + A'$ or $B + C$,
each of them having $|G_v| = 12$ and forming a single orbit with length $\ell = |G|/|G_v| = 6$.

It is also remarkable to note that the $\Z_4$ subgroup of $G$
generated by $d$ is {\em always} broken to $\Z_2$.
At first sight, it defies the intuition developed with the $D_4$ and $S_4$ cases,
where the $\Z_4$ subgroup could be broken or could survive.
Of course, one can write down vev alignments invariant under $d$,
but it happens that the additional structures in the $\Sigma(36)$ potential
preclude them from being a viable global minimum.
\end{itemize}

\section{Discussion}\label{section-summary}

\subsection{The overall picture}

\begin{table}[htb]
\centering
  \begin{tabular}{| r c | c c | c |}
\hline
group & $|G|$ & $|G_v|_{min}$ & $|G_v|_{max}$ & sCPv possible? \\   \hline\hline
abelian & $2,\, 3,\, 4,\ 8$  & 1 & $|G|$ & yes \\
$\Z_3 \rtimes \Z_2^*$ & 6 & 1 & 6  & yes \\
$S_3$ & 6 & 1 & 6  & ---\\
\hline
$\Z_4 \rtimes \Z_2^*$ & 8 & 2 & 8  & no\\
$S_3\times\Z_2^{*\dag}$ & 12 & 2 & 12  & yes\\
$D_4\times\Z_2^{*\dag}$ & 16 & 2 & 16  & no\\
\hline
$A_4\rtimes\Z_2^{*\dag}$ & 24 & 4 & 8  & no\\
$S_4\times \Z_2^*$ & 48 & 6 & 16  & no\\
$CP$-violating $\Delta(54)$ & 18 & 6 & 6  & ---\\
$CP$-conserving $\Delta(54)$ & 36 & 6 & 12 & yes \\
$\Sigma(36)$ & 72 & 12 & 12 & no\\
\hline
\end{tabular}
\caption{The amount of residual symmetry possible after EWSB for each discrete symmetry group of the 3HDM scalar potential 
(see text for details).
$\Z_2^*$ denotes either the usual CP symmetry or a GCP transformation 
(marked with $\dag$); 
see the main text for details on each group.
}
\label{table2}\end{table}

Let us now bring together all the results obtained in form of a list 
of the strongest and weakest breaking possible for each discrete symmetry in 3HDM.
This list is presented in Table~\ref{table2}.
The maximal amount of symmetry breaking of a given group $G$ 
corresponds to the smallest residual symmetry group  $G_v$, whose order is denoted by $|G_v|_{min}$,
while the minimal breaking corresponds to the largest residual symmetry group,
with order $|G_v|_{max}$.
The groups in the upper block, being suffiently small, allow for all types of symmetry breaking:
complete, partial, or no breaking at all.
The groups in the middle block can remain intact at the global minimum, but if they are broken,
their breaking is only partial. 
The last block contains groups which can neither remain unbroken nor break completely.
They are always broken to a proper subgroup.
Thus, for sufficiently large groups, only option (ii) mentioned in Section~\ref{sec:break.large} is available.

For symmetry groups with explicit $CP$-conservation, it is indicated in last column of Table~\ref{table2}
whether spontaneous $CP$-violation can occur upon minimization of the potential.
It is curious to note that explicit and spontaneous $CP$-violations always {\em come in pairs}.
Spontaneous $CP$-violation of a $G$-symmetric 3HDM can happen
only for those groups $G$, for which there exists an explicitly $CP$-violating counterpart.
If a Higgs-family symmetry protects the 3HDM from explicit $CP$-violation,
it also protects it from spontaneous $CP$-violation.
Whether this is just a coincidence or reveals a generic fact in NHDM is not yet known.
The general pattern, however, remains: for suffiently large 
symmetries containing accidental CP, the latter cannot be broken spontaneously.
As a related example, supersymmetric multi-Higgs extensions of the SM cannot break 
CP spontaneously\,\cite{scpv:susy}.

\subsection{Towards the $N$-doublet case}

The results obtained in 3HDM on the basis of explicit calculations can provide hints
at discrete symmetry breaking properties in NHDM, with general $N$.
We already noticed above that explicit and spontaneous $CP$-violation seems to come in pairs,
and it would be interesting to check whether this feature is generic.

Another observation made above is that sufficiently large discrete symmetries
must be broken partially. Although ``sufficiently large'' is a vague term,
the tendency itself has solid algebraic and group-theoretic grounds.
Indeed, suppose we work in NHDM with $N$ Higgs doublets in an irrep of $G$.
Then the only bilinear term in the potential compatible with this symmetry is $\sum_i \phi_i^\dagger \phi_i$.
In this case, there exists no neutral vacuum conserving this symmetry for $N \ge 2$, and 
no vacuum at all, including the charge-breaking one, for $N > 2$. 
Such symmetry groups are always  broken upon EWSB.

On the other hand, large symmetry breaking cannot be arbitrarily strong.
Indeed, if the group $G$ breaks down to a small subgroup $G_v$, 
then there must be at least $|G|/|G_v|$ separate degenerate global minima in the orbit space.
However, as explained for 2HDM in \cite{Ivanov2HDMMinkowski} and adapted for NHDM in \cite{IvanovNishiNHDM,IvanovNHDM-II},
search for the global minimum of the Higgs potential can be cast in pure geometric terms.
In this picture, the global minima arise as the contact points of two algebraic manifolds of certain degrees
defined in the real space $\RR^{N^2}$. 
The number of such contact points must be bounded from above by some sort of multi-dimensional 
generalization of the B\'ezout's theorem. Let us denote the maximum number of
such contact points for NHDM as $p_N$.
Then, if $|G|> p_N$, the group $G$ cannot break completely.

For 2HDM, $p_2 = 2$, and for 3HDM, as suggested by the present work, $p_3 \ge 8$.
For general $N$, the exact value of $p_N$ is unknown.
Developing the algebraic-geometric methods to the point when $p_N$ can be found,
would constitute a significant step forward in understanding symmetry breaking patterns in their general set up.

The above geometric reinterpretation of the minimization problem also makes it very plausible
that symmetry breaking patterns strongly depend on the algebraic degree of the potential.
Indeed, the usual renormalizable Higgs potential can be represented as a quadric (degree-2 algebraic manifold)
in the space of $N^2$ real bilinears. Adding sextic terms makes it a degree-3 manifold.
Since B\'ezout's theorem for intersection of planar curves explicitly depends on their degrees,
the same can readily be expected for its conjectured higher-dimensional generalization. Thus,
$G$-symmetric potentials with sextic terms can have {\em more} degenerate minima
than a renormalizable potential with the same symmetry group $G$.
The increased amount of global minima opens a possibility for {\em stronger} symmetry breaking
than what was possible only with quadratic and quartic potentials.
It would be very interesting to build an explicit realizable of this possibility and 
investigate its phenomenological consequences.
\\

To summarize, in this paper, we systematically investigated how discrete symmetry groups $G$ of the 3HDM 
scalar potential break upon minimization of the potential.
We checked one by one all $G$'s allowed in 3HDM and all vev alignment which can arise for each $G$,
and listed the residual symmetry groups $G_v$. 
Table~\ref{table2} summarizes the strongest and weakest symmetry breaking for each $G$,
as well as the possibility of spontaneous $CP$-violation.
These result led us to a number of observations, which might hold for more than three Higgs doublets
and, perhaps, for more general Higgs sectors.
However, checking them will require yet additional algebraic-geometric or group-theoretic
tools.
\\

We are thankful to Jo\~ao Silva and Enrico Nardi for useful discussions and comments.
The work of C.C.N. was partially supported by Brazilian CNPq and Fapesp 
(2013/26371-5 and 2013/22079-8).

\appendix

\section{Absence of complete symmetry breaking in $D_4$ and $CP$-conserving $S_3$ 3HDM}\label{appendix-D4}

Here we prove that symmetry groups $G = D_4 \times \Z_2^*$ or $S_3 \times \Z_2^*$ cannot be broken completely 
in 3HDM via minimization of a renormalizable potential.
To show it, we will evaluated the Higgs potential at the classical field values $(v_1,\, v_2 e^{i\xi_2},\, v_3e^{i\xi_3})$
with non-zero $v_i$ and will find that extremization either (1) sets the phases $\xi_i$ to zero, 
up to a rephasing transformation from $G$, or (2) sets $v_2 = v_3$ and $\xi_3= \xi_2$ or $ - \xi_2$. 
In either case, the extremum remains invariant under a residual symmetry from group $G$. 

We find it instructive to start with the $D_4$ case. Although we already know that the presence of $\Z_4$
subgroup implies $CP$ conservation for the potential and a GCP symmetry in the vacuum,
we will rederive it in another way to demonstrate a technique to be used for the $S_3 \times \Z_2^*$ model.

{\bf $D_4$ 3HDM.}
As usual, we write the most general $D_4$-symmetric 
potential as $V_{D_4} = V_0 + V_{ph}$, with $V_0$ given by (\ref{V0new}) and 
the phase-dependent part written as
\be
V_{ph.} = {1\over 2}\lambda_5 \left[(\phi_1^\dagger \phi_2)^2 + (\phi_1^\dagger \phi_3)^2\right] + 
{1\over 2}\lambda_6 (\phi_2^\dagger \phi_3)^2 + h.c.
\ee
where all parameters are real.
Since $D_4 \times \Z_2^*$ contains as subgroups two groups we studied previously, namely 
$\Z_2 \times \Z_2 \times \Z_2^*$ and $\Z_4 \rtimes \Z_2^*$, we could write $V_{ph}$ 
in the form of (\ref{VphZ2Z2Z2}) or (\ref{VphZ4}).
Our choice is based on (\ref{VphZ2Z2Z2}) and it differs from (\ref{VphZ4}) by a basis change.
The symmetry group of $V_{D_4}$ is generated by independent sign flips, by $c$, and by the $CP$ transformation.

Positive coefficients $\lambda'_3$ and $\lambda'_4$ in (\ref{V0new}) guarantee that the minimum is neutral.
As usual, we write $\lr{\phi_i^0} = v_i e^{i\xi_i}/\sqrt{2}$, which makes $\lr{z_{ij}} = 0$.
Now, we also define three 2D vectors
\be
\vec{r}_i \equiv {v_i^2 \over 2}(\cos2\xi_i,\, \sin2\xi_i),
\ee
so that
\be
(\phi_i^\dagger \phi_i) = r_i \equiv |\vec r_i|\,,\quad 
{1\over 2}\left[(\phi_i^\dagger \phi_j)^2 + (\phi_j^\dagger \phi_i)^2\right] = 
\vec r_i\cdot \vec r_j\,. 
\ee
With this notation, the full $D_4$-symmetric potential takes the following form:
\bea
V &=& - m_1^2 r_1 - m_2^2 (r_2+r_3) + {1 \over 2}\lambda_1 r_1^2 + {1 \over 2}\lambda_2 (r_2^2 + r_3^2)
+ \lambda_3 r_1(r_2+r_3) + \lambda_4 r_2r_3 \nonumber\\
&&+ \lambda_5 (\vec r_1\cdot \vec r_2 + \vec r_1\cdot \vec r_3) + \lambda_6 \vec 
r_2\cdot \vec r_3\,.
\label{D4viari}
\eea
In order to check whether the group can be broken completely, we search for extrema with non-zero $v_i$.
The extremization problem can then be formulated in terms of gradients: $\vec\nabla_{i} V = 0$.
Recalling that $\vec\nabla r^2 = 2\vec r$ and $\vec \nabla r = \vec r/r$, we obtain:
\bea
&&\left[\lambda_1 r_1 + \lambda_3 (r_2+r_3)\right]\vec r_1 + \lambda_5 r_1(\vec r_2 + \vec r_3) = m_1^2 \vec r_1\,,\label{D4nabla1}\\
&&\left[\lambda_2 r_2 + \lambda_3 r_1 + \lambda_4 r_3\right]\vec r_2 + \lambda_5 r_2 \vec r_1 + \lambda_6 r_2 \vec r_3 = m_2^2 \vec r_2\,,\label{D4nabla2}\\
&&\left[\lambda_2 r_3 + \lambda_3 r_1 + \lambda_4 r_2\right]\vec r_3 + \lambda_5 r_3 \vec r_1 + \lambda_6 r_3 \vec r_2 = m_2^2 \vec r_3\,.\label{D4nabla3}
\eea
Note that these are vectorial equalities.
From (\ref{D4nabla1}) we conclude that either $\lambda_5 = 0$ or $\vec r_2 + \vec r_3$ is parallel to $\vec r_1$. 
We exclude the former choice because it leads to continuous symmetries in the potential, so we write 
\be
\vec r_1 = - c (\vec r_2 + \vec r_3) \label{r1r2r3}
\ee
with the coefficient $c$ to be determined. Using this relation, we replace $\vec r_1$ in (\ref{D4nabla2}) and (\ref{D4nabla3}) and obtain:
\bea
&&\left[\lambda_2 r_2 + \lambda_3 r_1 + \lambda_4 r_3 - \lambda_5 c r_2\right]\vec r_2 + (\lambda_6 - c\lambda_5)r_2 \vec r_3 = m_2^2 \vec r_2\,,\label{D4nabla4}\\
&&\left[\lambda_2 r_3 + \lambda_3 r_1 + \lambda_4 r_2 - \lambda_5 c r_3\right]\vec r_3 + (\lambda_6 - c\lambda_5)r_3 \vec r_2 = m_2^2 \vec r_3\,.\label{D4nabla5}
\eea

We have here two options: either all three vectors $\vec r_i$ are aligned (case A) or not (case B). 
In case B we then have $c = \lambda_6/\lambda_5$.

In case A, the alignment of $\vec r_i$ means that the phases $\xi_{2}$ and $\xi_3$ are multiples of $\pi/2$.
Even if all three vevs are different, these phases lead to a residual symmetry:
in the case $(v_1,\, v_2,\, v_3)$ it is just $CP$, in the case $(v_1,\, v_2,\, i v_3)$, it is the GCP transformation $\sigma_{23}\cdot CP$.

In case B, non-trivial phases are still allowed. Equations (\ref{D4nabla4}) and (\ref{D4nabla5}) 
can be simplified as equations on coefficients in front of $\vec r_2$ and $\vec r_3$ and lead to
\be
(\lambda_2 - \lambda_6) r_2 + \lambda_3 r_1 + \lambda_4 r_3 = m_2^2\,,
\quad 
(\lambda_2 - \lambda_6) r_3 + \lambda_3 r_1 + \lambda_4 r_2 = m_2^2\,.
\ee
Their difference leads to 
\be
(\lambda_2 - \lambda_4 - \lambda_6)(r_2 - r_3) = 0\,.
\ee
Once again, we have two options. If $\lambda_2 - \lambda_4 - \lambda_6 = 0$,
the potential acquires a flat direction because it now depends
only on $r_1$, $r_2+r_3$, and $\vec r_2+\vec r_3$, but not on $r_2 - r_3$. This means that 
there is a continuum of global minima (an ellipse) with the same $r_1$, $r_2+r_3$, and $\vec r_2+\vec r_3$,
but different $r_2 - r_3$. We disregard this situation. 
The only remaining possibility is to set $r_2 = r_3$. The vev alignment is now of type
\be
(v_1,\, \pm v_2e^{i\xi},\, \pm v_2 e^{-i\xi})\,,
\ee
where $\pm$ signs are independent. This alignment also possesses a residual symmetry:
$c\cdot CP$ times sign flips when necessary.
Thus, in either case, the full symmetry group $G = D_4 \times \Z_2^*$ is broken not completely
but down to the $\Z_2^*$ group generated by a GCP transformation.\\

{\bf $CP$-conserving $S_3$ 3HDM.}
We will now apply the same method to the $S_3\times \Z_2^*$ 3HDM.
We use the same $V_0$ and the phase-dependent part $V_{ph}$ in the form (\ref{VphS3})
with real $\lambda_5$ and $\lambda_6$.
Now, we introduce another set of 2D vectors
\be
\vec s_i = {\sqrt{2} \over v_i} (\cos\alpha_i,\, \sin\alpha_i)\,, 
\ee
where
\be
\alpha_1 = \xi_2 - \xi_3\,,\quad \alpha_2 = \xi_3 - \xi_1\,,\quad \alpha_3 = \xi_1 - \xi_2\,.
\ee
Then $V_0$ takes the same form as in the first line of (\ref{D4viari}) with $r_i = 1/{\vec s_i}^{\, 2}$, while
$V_{ph}$ can be written as
\be
V_{ph} = r_1 r_2 r_3\left[\lambda_5 \vec s_2\cdot \vec s_3 + \lambda_6 \vec 
s_1\cdot(\vec s_2 + \vec s_3)\right]\,.
\label{S3viasi}
\ee
Thus, the potential is now written in terms of three vectors $\vec s_i$.
We can again cast the extremization problem in terms of conditions $\vec \nabla_i V = 0$.
The calculations become a bit more cumbersome, powers of $r_i$ floating around,
but we nevertheless encounter the same options: 
either all $\vec s_i$ are aligned, or $\vec s_2 + \vec s_3$ is parallel to $\vec s_1$ and $v_2 = v_3$.
In terms of phases $\xi_i$, the former case leads to real vevs $(v_1,\, v_2,\, 
v_3)$, up to a rephasing by $a_3$,
while the latter case produces alignment $(v_1,\, v_2e^{i\xi_2},\, v_2 e^{i\xi_2})$.
In either case we get a residual symmetry in the vacuum: 
either a GCP symmetry, or the $\phi_2 \leftrightarrow \phi_3$ symmetry.
Thus, $S_3 \times \Z_2^*$ is broken to $\Z_2$ or $\Z_2^*$.

Finally, the existence of a solution with $v_2 \not = v_3$ in $CP$-conserving $S_3$ 3HDM
means that such a solution is also possible for its $CP$-violating version.
In this case, phases are irrelevant, and the entire symmetry group $S_3$ is broken.

\section{Global minima of the $\Delta(27)$-family potentials}\label{appendix-D27}

Here, we use the geometric method of \cite{minimization2012}
to find all possible vev alignments 
for the $\Delta(27)$-family of symmetry groups in 3HDM.

The first step is to construct the orbit space in terms of suitable variables.
Let us introduce the following real quantities:
\bea
r_0 &=& {1\over \sqrt{3}}\left(\phi_1^\dagger \phi_1+ \phi_2^\dagger \phi_2+\phi_3^\dagger \phi_3\right)\,,\nonumber\\
X &=& 
 \biggl\{{1 \over 3}\left[(\phi_1^\dagger \phi_1)^2+ (\phi_2^\dagger \phi_2)^2+(\phi_3^\dagger \phi_3)^2
- (\phi_1^\dagger \phi_1)(\phi_2^\dagger \phi_2) - (\phi_2^\dagger \phi_2)(\phi_3^\dagger \phi_3) 
- (\phi_3^\dagger \phi_3)(\phi_1^\dagger \phi_1)\right] \nonumber\\
&&\qquad \qquad + \ |\phi_1^\dagger \phi_2|^2 + |\phi_2^\dagger \phi_3|^2 + |\phi_3^\dagger \phi_1|^2\biggr\}\,, \nonumber\\
X' &=& |\phi_1^\dagger \phi_2|^2 + |\phi_2^\dagger \phi_3|^2 + |\phi_3^\dagger \phi_1|^2\,,\nonumber \\
Y &=& {1 \over 3}\left[|\phi_1^\dagger \phi_2 - \phi_2^\dagger \phi_3|^2 + |\phi_2^\dagger \phi_3 - \phi_3^\dagger \phi_1|^2 + 
|\phi_3^\dagger \phi_1 - \phi_1^\dagger \phi_2|^2\right]\,,\\
Y' &=& {2\over \sqrt{3}} \Im  \left[(\phi_1^\dagger \phi_2)(\phi_1^\dagger \phi_3) + (\phi_2^\dagger \phi_3)(\phi_2^\dagger \phi_1) + (\phi_3^\dagger 
\phi_1)(\phi_3^\dagger \phi_2)\right]\,,
\eea
as well as the corresponding rescaled variables
\be
x = X/r_0^2\,,\quad x' = X'/r_0^2\,,\quad y = Y/r_0^2\,,\quad y' = Y'/r_0^2\,.
\ee
The potential (\ref{potential-Delta54}) can be written as a linear combination of these quantities:
\be
V_1 = - M^2 r_0 + r_0^2\left(\Lambda_0 + \Lambda_1 x + \Lambda_1' x' + \Lambda_2 y + \Lambda_2' y '\right)\,,
\label{V1linear}
\ee
where $M^2 = \sqrt{3} m^2$, $\Lambda_0 = 3\lambda_0$, $\Lambda_1 = \lambda_1$,
$\Lambda_1' = \lambda_2 - \lambda_1 + 2 \Re \lambda_3$,
$\Lambda_2 = - 3 \Re \lambda_3$, and
$\Lambda_2' = -\sqrt{3}\Im \lambda_3$.
It is known that $1/4 \le x \le 1$ in 3HDM \cite{IvanovNishiNHDM}, 
and the neutral vevs correspond to $x=1$. Setting $x=1$ in (\ref{V1linear}), 
we rewrite the potential as a linear function defined in the 3D space $(x',\, y,\, y')$.

\begin{figure}[!htb]
   \centering
\includegraphics[width=8cm]{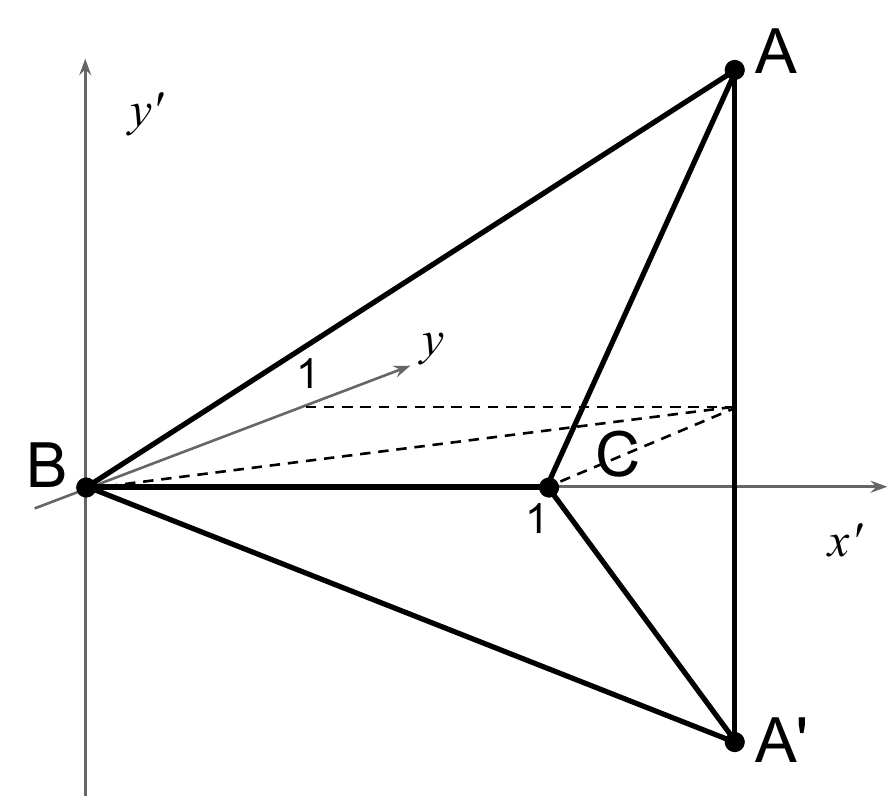}
\caption{The tetrahedron in the $(x',y,y')$ space for the $\Delta(27)$-family of symmetries}
   \label{fig-tetrahedron}
\end{figure}

Next, we find inequalities these variables satisfy.
From definitions, we have $0 \le x' \le 1$ and $0 \le y \le x'$, 
where the last inequality comes from 
\be
X'-Y = {|(\phi_1^\dagger \phi_2) + (\phi_2^\dagger \phi_3) + (\phi_3^\dagger \phi_1)|^2 \over 3} \ge 0\,.\label{X'-Y}
\ee
In addition, we notice that 
\bea
Y - Y' &=& 
{2\over 3}\biggl\{|\phi_1^\dagger \phi_2|^2 + |\phi_2^\dagger \phi_3|^2 + |\phi_3^\dagger \phi_1|^2  
+ 2\Re \left[\omega (\phi_1^\dagger \phi_2)(\phi_1^\dagger \phi_3) + 
    \omega (\phi_2^\dagger \phi_3)(\phi_2^\dagger \phi_1) + \omega (\phi_3^\dagger \phi_1)(\phi_3^\dagger \phi_2)\right]\biggr\} \nonumber\\
&=& {2\over 3}\left|\phi_1^\dagger \phi_2 + \phi_2^\dagger (\omega \phi_3) + (\omega\phi_3)^\dagger \phi_1\right|^2  \ge 0\,,
\eea
and similarly for $Y + Y'$:
\be
Y+Y'= {2\over 3}\left|\phi_1^\dagger \phi_2 + \phi_2^\dagger (\omega^2 \phi_3) + (\omega^2\phi_3)^\dagger \phi_1\right|^2  \ge 0\,.
\ee
So, summarizing all restrictions, we have:
\be
0 \le x' \le 1\,,\quad 0 \le y \le x'\,,\quad |y'| \le y\,.\label{restrictions}
\ee
These inequalities define a tetrahedron in the $(x',y,y')$ space shown in Fig.~\ref{fig-tetrahedron}.
The orbit space must lie inside or on the boundaries of this tetrahedron.,
but it does not have to fills it completely.

The next step is to make sure that the four vertices of this tetrahedron, labeled in Fig.~\ref{fig-tetrahedron} by $A$, $A'$, $B$, and $C$,
{\em do belong} to the orbit space.
Here is the explicit derivation.
\begin{itemize}
\item
Point $A$ is at $(x',y,y')=(1,1,1)$. From $x-x'=0$ we deduce that $|v_1|=|v_2|=|v_3|$. From (\ref{X'-Y}) we deduce that
$(\phi_1^\dagger \phi_2) + (\phi_2^\dagger \phi_3) + (\phi_3^\dagger \phi_1) = 0$. These two conditions, together with the positive sign
of $y'$, are satisfied only by the following three vev alignments $(v_1,\, v_2,\, v_3)$:
\be
\mbox{point $A$:}\quad (1,\,1,\,\omega)\,,\quad (1,\,\omega,\,1)\,,\quad(\omega,\,1,\,1)\,.
\ee
\item
Point $A'$ is at $(x',y,y')=(1,1,-1)$.  The conditions are the same but $y'$ is now negative, which is possible only at
\be
\mbox{point $A'$:}\quad (1,\,1,\,\omega^2)\,,\quad (1,\,\omega^2,\,1)\,,\quad(\omega^2,\,1,\,1)\,.
\ee
\item
Point $C$ is at $(x',y,y')=(1,0,0)$. Again, we have $|v_1|=|v_2|=|v_3|$ plus certain conditions on phases, which can be all satisfied only at
\be
\mbox{point $C$:}\quad (1,\,1,\,1)\,,\quad (1,\,\omega,\,\omega^2)\,,\quad(1,\,\omega^2,\,\omega)\,.
\ee
\item
Point $B$ is at $(x',y,y')=(0,0,0)$, which is possible only at 
\be
\mbox{point $B$:}\quad (1,\,0,\,0)\,,\quad (0,\,1,\,0)\,,\quad(0,\,0,\,1)\,.
\ee
\end{itemize}

The key statement now is that the four points $A$, $A'$, $B$, and $C$ are {\em the only options where the global minimum 
can be}, provided we require that there be no massless
physical Higgs bosons.

\begin{figure}[!htb]
   \centering
\includegraphics[width=8cm]{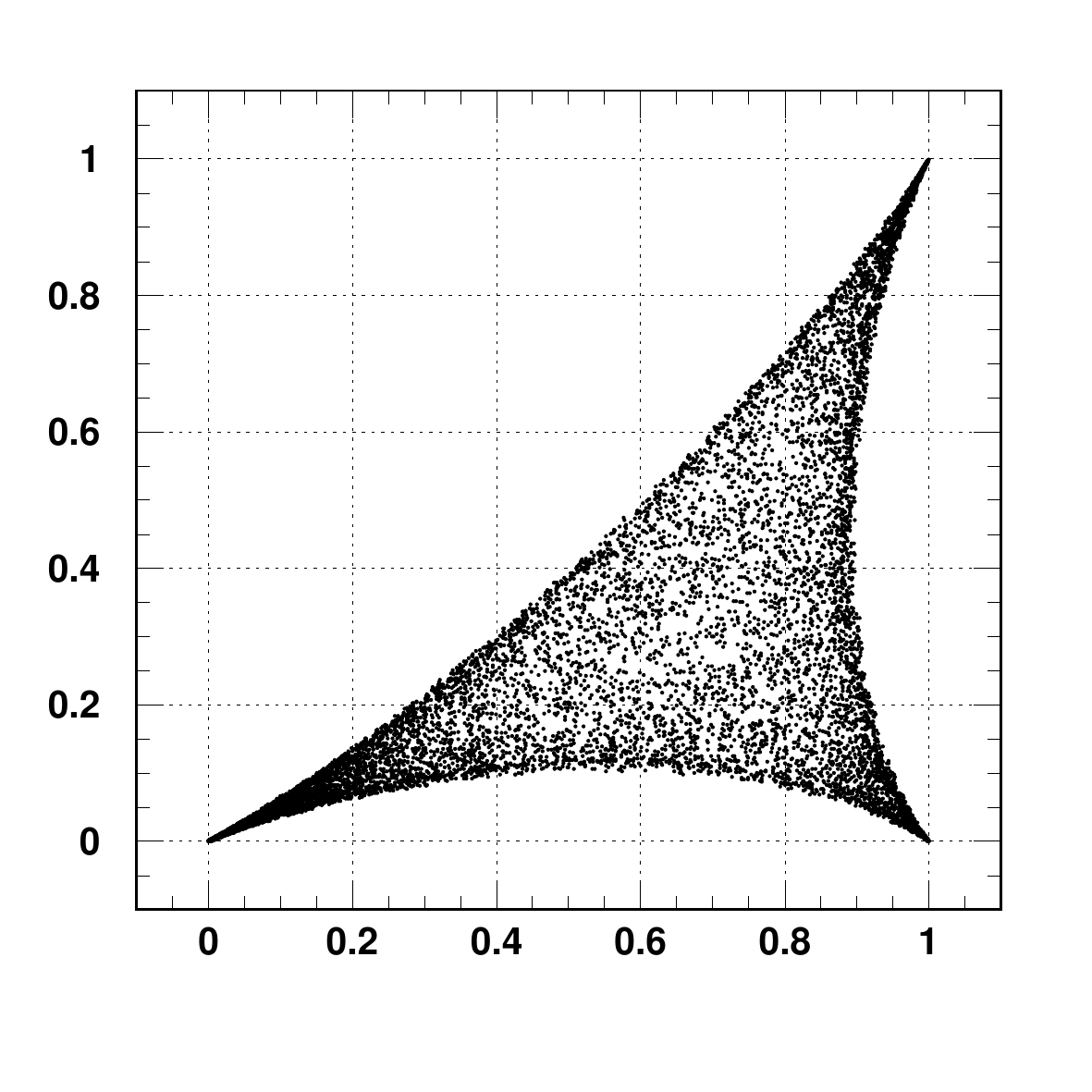}\hfill
\includegraphics[width=8cm]{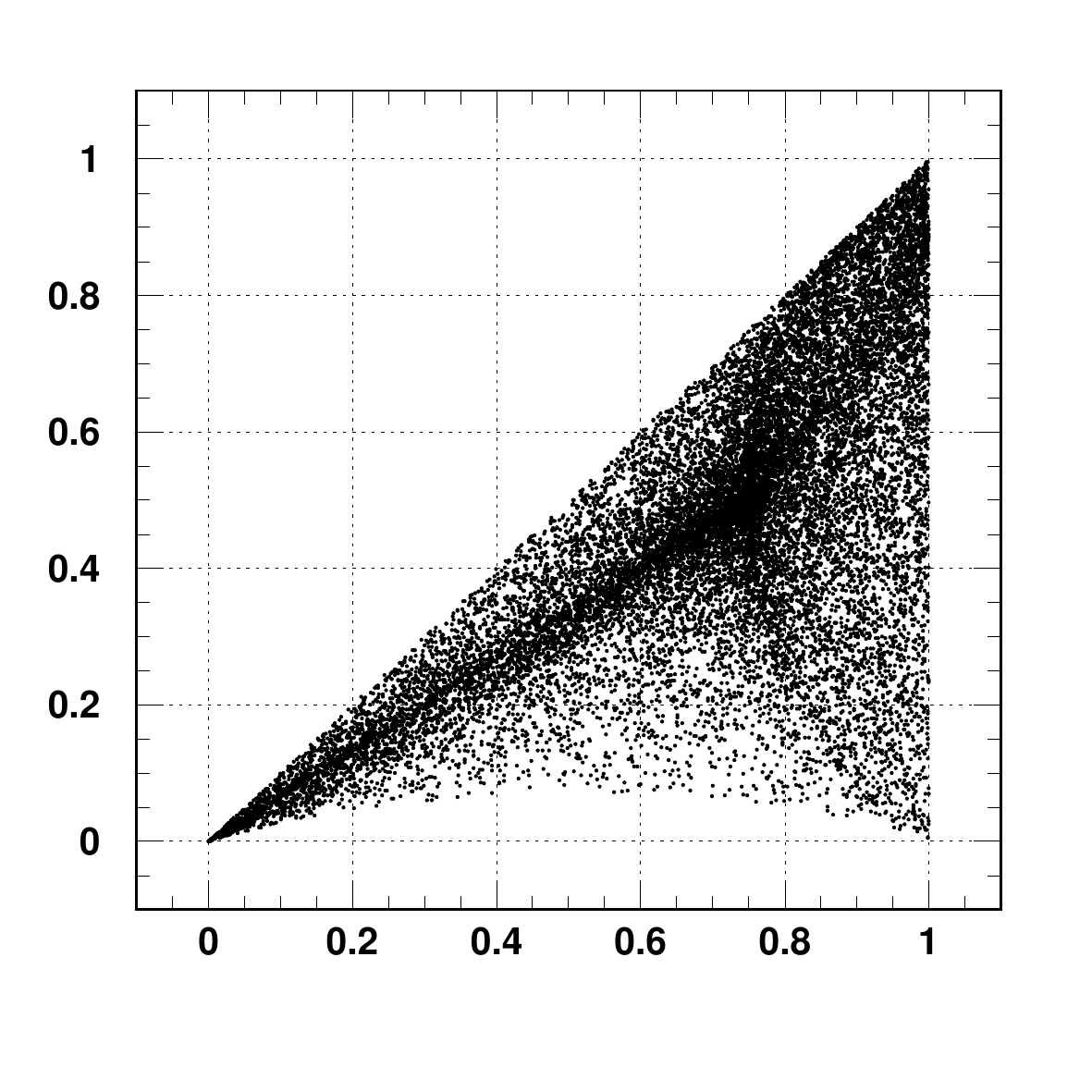}
\caption{Left: The projection of one face of the $\Delta(54)$ orbit space on the $(x',y)$ plane; 
shown are the points with $y-y' \le 0.001$ out of $10^7$ points randomly selected from the neutral orbit space.
Right: The orbit space in the $(x',y)$ space for $\Delta(54)\rtimes \Z_2$ (random scan with $20,000$ points). }
   \label{fig-astroidlike}
\end{figure}

The proof follows from the geometric methods of \cite{minimization2012}
and also resembles what is known as linear programming in mathematics.
Since all four vertices of the tetrahedron belong to the orbit space,
there can be no global minimum lying strictly inside the tetrahedron for any combination of the free parameters $\Lambda_i$, 
What remains to be checked is whether there are additional isolated points of the orbit
space lying on the edges or faces of the tetrahedron.
Although this should be doable algebraically, we use here a numerical shortcut.
Namely, we scan the orbit space by randomly choosing the three complex $v_i$'s,
calculating the corresponding $(x',y,y')$ points, and then checking very thin slices lying at the faces.
Fig.~\ref{fig-astroidlike}, left, shows the results of this exercise for the $ABC$ face of the tetrahedron;
other faces lead to similar results.
One sees that the points densely cover an astroidlike shape. There are no points lying on the edges,
and there are no isolated points lying on the face.
This means that if a point on a face happens to be a global minimum, then {\em the entire face}
will also correspond to the global minimum, and this implies massless physical Higgses.

The above construction describes the orbit space of the $CP$-violating $\Delta(54)$ 3HDM, 
the minimal realizable symmetry from the $\Delta(27)$-family.
In what concerns higher symmetry groups from this family,
they are obtained by the simple projection of the entire construction from the $(x',y,y')$ space
onto subspaces.
Namely, the orbit space of the $CP$-conserving $\Delta(54)$ model is the projection on the $y'=0$ plane,
and the entire neutral orbit space has the shape shown in Fig.~\ref{fig-astroidlike}, right.
For the $\Sigma(36)$-symmetric 3HDM,
we set $\Lambda_1'$ to zero, and the potential does not depend on $x'$.
The orbit space is then obtained by further projecting the shape of Fig.~\ref{fig-astroidlike}, right,
onto the $y$ axis (the vertical line),
and is represented by the line segment $0 \le y \le 1$.

These two projections satisfy the following properties:
they map vertices to vertices,
and they {\em do not map anything else} to vertices.
Therefore, the global minima in these cases are the same points $A$, $A'$, $B$, and $C$,
some of them merged, but nothing extra. 
The $CP$-conserving $\Delta(54)$ model can have global minima at points $B$, $C$,
or $A$ and $A'$ taken simultaneously.
The $\Sigma(36)$ model can have only two kinds of global minima: either $B$ and $C$
simultaneously, or $A$ and $A'$ simultaneously.

\end{document}